\documentclass[trackchanges, twocolumn]{aastex701}
\usepackage{xcolor}
\definecolor{linkcolor}{rgb}{0.0,0.3,0.5}

\definecolor{amethyst}{rgb}{0.8, 0.0, 0.0}



\usepackage{graphicx}
\usepackage{amsmath}	
\usepackage{subcaption}

\usepackage{textgreek}
\usepackage{tensind}
\tensordelimiter{?}

\usepackage{orcidlink}
\usepackage{soul}  
\usepackage{cancel}
\usepackage{multirow}
\usepackage{comment}
\usepackage[normalem]{ulem}

\usepackage{xcolor}
\newcommand{\rev}[1]{#1}

\begin{document}

\title{\large The abundance and emission of PAHs from the local to early Universe}

\author[orcid=0000-0002-9729-3721,sname='Max Parente']{Massimiliano Parente}
\affiliation{Department of Astronomy, University of Florida, 211 Bryant Space Sciences Center, Gainesville, FL 32611, USA}
\email[show]{parente.m@ufl.edu}  

\author[orcid=0000-0002-7064-4309]{Desika Narayanan}
\affiliation{Department of Astronomy, University of Florida, 211 Bryant Space Sciences Center, Gainesville, FL 32611, USA}
\affiliation{Cosmic Dawn Center at the Niels Bohr Institute, University of Copenhagen and DTU-Space, \\ Technical University of Denmark, Denmark}  
\email[]{desika.narayanan@ufl.edu}

\author[orcid=0000-0002-5653-0786]{Paul Torrey}
\affiliation{Department of Astronomy, University of Virginia, 530 McCormick Road, Charlottesville, VA 22903, USA}
\affiliation{Virginia Institute for Theoretical Astronomy, University of Virginia, Charlottesville, VA 22904, USA}
\affiliation{The NSF-Simons AI Institute for Cosmic Origins, USA}
\email[]{}

\author[orcid=0000-0002-4480-6909]{Gian Luigi Granato}
\affiliation{INAF, Osservatorio Astronomico di Trieste, via Tiepolo 11, I-34131, Trieste, Italy}
\affiliation{IFPU, Institute for Fundamental Physics of the Universe, Via Beirut 2, 34014 Trieste, Italy}
\email[]{}

\author[orcid=0000-0001-7449-4638]{Brandon S. Hensley}
\affiliation{Jet Propulsion Laboratory, California Institute of Technology, 4800 Oak Grove Drive, Pasadena, CA 91109, USA}
\email[]{}

\author[orcid=0000-0002-6149-8178]{
Jed McKinney}
\affiliation{Department of Astronomy, University of Texas at Austin, Austin, TX, USA}
\email[]{}

\author[orcid=0000-0001-8592-2706]{
Alexandra Pope}
\affiliation{Department of Astronomy, University of Massachusetts, Amherst, MA, 01003, USA}
\email[]{}

\author[orcid=0000-0003-1151-4659]{
Gergö Popping}
\affiliation{European Southern Observatory, Karl-Schwarzschild-Str. 2, 85748, Garching, Germany}
\email[]{}

\author[orcid=0000-0003-4702-7561]{
Irene Shivaei}
\affiliation{Centro de Astrobiologia (CAB), CSIC-INTA, Carretera de Ajalvir km 4, Torrejon de Ardoz, 28850, Madrid, Spain}
\email[]{}

\author[orcid=0000-0002-7571-5217]{Laura Silva}
\affiliation{INAF, Osservatorio Astronomico di Trieste, via Tiepolo 11, I-34131, Trieste, Italy}
\affiliation{IFPU, Institute for Fundamental Physics of the Universe, Via Beirut 2, 34014 Trieste, Italy}
\email[]{}

\author[orcid=0000-0003-1545-5078]{
J.D.T. Smith}
\affiliation{Ritter Astrophysical Research Center, Department of Physics and Astronomy, University of Toledo, Toledo, OH 43606, USA}
\email[]{}

\begin{abstract}

We present the first large-volume cosmological simulation of galaxy evolution with a model for the abundance and luminous emission of Polycyclic Aromatic Hydrocarbons (PAHs) across a wide range of cosmic time ($0 < z \lesssim 6$). The simulation is built on a semi-analytic model (SAM) that includes a treatment of dust size distribution, with PAHs being identified with the smallest carbonaceous grains ($< 15\,\text{\AA}$) residing in the diffuse ISM of galaxies. PAH emission is computed by post-processing the simulation output with the radiative transfer code GRASIL.
In our model, \rev{where stars only inject large grains,} the PAH abundance emerges from the growth \rev{and} shattering of dust grains in the ISM, naturally reproducing the observed correlation between the PAH mass fraction ($q_{\rm PAH}$) and both metallicity and specific star formation rate (sSFR).  We derive a relation to infer $q_{\rm PAH}$ from $L_{\rm PAH}$, $L_{\rm IR}$, and sSFR, with a relatively small scatter (a factor of $\approx 1.3$).
Simulated PAH emission correlates well with key galaxy properties such as SFR and molecular gas content, and reproduces observational constraints.  Taken together, these results support the viability of a top-down formation scenario for PAHs.
We extend our model predictions to higher redshifts, which can be exploited for future missions such as PRIMA. This work is a significant step toward leveraging PAH observations as a tool for studying galaxy evolution.

\end{abstract}



\section{Introduction}
It has been about half a century since the mid-infrared (MIR; $\lambda \approx 3-20 \, \mu{\rm m}$) spectra of some astrophysical objects were  found to be dominated by strong emission features -- now recognized to include prominent bands at 3.3, 6.2, 7.7, 8.6, 11.3, 12.7, and 17 $\mu$m -- as first reported by \cite{Gillett73} in ionized nebulae. Approximately a decade later, these ``\textit{unidentified infrared bands}'' were attributed to the vibrational de-excitation of Polycyclic Aromatic Hydrocarbon (PAH) molecules by \citet{Leger84} and \citet{Allamandola85}.  ISO and Spitzer/IRS spectroscopy subsequently found that these features are common in star-forming galaxies, a result recently extended to cosmic noon ($z\sim2$) and beyond with JWST/MIRI \citep{Shivaei24}. 
PAHs play a central part in the physics of the interstellar medium (ISM) itself, e.g., by catalyzing the formation of molecular hydrogen on grain surfaces \citep[e.g.,][]{Thrower12, Foley18, curro26} and by dominating the photoelectric heating of the neutral gas \citep{Bakes94}.

PAHs are planar aromatic hydrocarbon molecules containing $\approx 20-1000$ carbon atoms \citep[e.g.,][]{Tielens08}. Their planar geometry is due to the strongly anisotropic bonding of carbon, which also allows them to host hydrogen atoms at the periphery of the ring cluster \citep{Leger84}. The resulting electronic and vibrational structure has been characterized extensively both in laboratory experiments and via quantum chemical calculations \citep[e.g.,][]{Bakes01a, Bakes01b, Malloci04}. A key difference with respect to larger dust grains is that a PAH molecule contains too few atoms to reach steady-state equilibrium with the ambient radiation field, so its emission cannot be treated with the standard equilibrium energy-balance approach. Instead, a PAH absorbs a single UV/optical photon, is heated to a peak temperature of order $10^3\,{\rm K}$, and cools by radiating in the characteristic aromatic C--C and C--H vibrational bands before the next photon arrives \citep{Leger84, Allamandola85, Puget89, DraineLi07}.

The origin of the PAH population in astronomical environments is still debated, with two broad families of models. In ``top-down'' scenarios, PAHs are produced by the collisional fragmentation of larger carbonaceous grains \citep[e.g.,][]{Jones96, Seok14}. In ``bottom-up'' scenarios, PAHs instead grow from smaller hydrocarbon precursors, either in the outflows of carbon-rich AGB stars \citep[e.g.,][]{Frenklach89, Cherchneff92} or, more generally, via hydrogen-abstraction/acetylene-addition (HACA) chemistry \citep{Reizer22} in the gas phase. Some recent models suggest both channels may operate in different regimes -- with PAH growth suppressed at low metallicities \citep{curro26} -- while the size distribution can additionally be reshaped by size-dependent photodissociation, which preferentially destroys the smallest PAHs \citep[e.g.,][]{Allain96, Montillaud13, Murga19}.

Despite the theoretical uncertainties related to PAH formation and survival in the ISM, PAH emission has been extensively used in both galactic and extragalactic astrophysics. PAH features and their relative intensities are widely used as a diagnostic of local ISM conditions, including the size and ionization state of the PAH population and the hardness/intensity of the interstellar radiation field \citep[e.g.,][]{DraineLi07, Maragkoudakis20, Draine21, Baron25}. Spatially resolved studies of photodissociation regions find systematic size and ionization variations that trace the local radiation field, generally suggesting preferential destruction of the smallest PAHs by UV photoprocessing \citep[e.g.,][]{Knight21, Chastenet19, Egorov23, Sutter24}, with JWST/MIRI now extending this kind of diagnostic to star forming galaxies at cosmic noon \citep{Donnan26, Lofaro26, Wang26, Kendrew26, Shivaei26}.
An important result found across these observations is the correlation between PAH emission\footnote{Or, alternatively, the PAH mass fraction, typically inferred by fitting the observed PAH features and dust continuum with a model spectral energy distribution.} and metallicity, with resolved studies consistently inferring a suppression of the PAH fraction toward lower local metallicity \citep[e.g.,][]{Sandstrom12, Chastenet19, Tarantino25, Whitcomb26}. This trend, known as the PAH-metallicity relation (PZR), has also been found on galaxy-integrated scales, from the local Universe \citep[e.g.,][]{Engelbracht05, Madden06, RR15, Aniano20} out to cosmic noon \citep{Shivaei24}. The physical origin of the PZR -- whether preferential UV destruction, delayed injection by AGB stars, or inhibited grain growth -- remains  debated \citep[e.g.,][]{Madden06, Galliano08, Xie19}.
At unresolved, galaxy-integrated scales, PAH features are also used as tracers of other physical quantities, with individual features (most commonly the $3.3,\,6.2, \,7.7,\,11.3$ $\mu$m bands) or their total luminosity calibrated across a wide range of galaxy types to trace the star formation rate \citep[SFR; e.g.,][]{Pope08, Shipley16, Xie19} and the molecular gas content \citep{Cortzen19} of galaxies, with these features now more easily accessible at cosmic noon thanks to JWST/MIRI \citep{Shivaei24, SB24, McKinney26}.

Given \textit{(i)} the empirically demonstrated power of PAH emission as a tracer of both local ISM conditions and galaxy-scale physical properties, \textit{(ii)} its prominence in the MIR spectra of star-forming galaxies, and \textit{(iii)} the growing observational interest driven by JWST capabilities, together with the prospect of pushing this frontier to even higher redshift with future infrared concepts such as PRIMA, there is significant interest recently in incorporating PAHs into the framework of galaxy evolution modeling, extending galaxy formation simulations so that they can predict, interpret, and help guide PAH observations across cosmic time.
While galaxy evolution simulations in the last decade have increasingly focused on modeling dust physics and emission \citetext{e.g., \citealp{Popping17, McKinnon18, Aoyama18, Li2019, Graziani20, Granato21, Parente22, Triani23, Dubois24, Yates24, Caleb_prep, Trayford25}; see review by \citealp{Parente25rev}}, the modeling of PAHs and the consequent MIR spectral energy distribution (SED) within this framework remains a relatively new field. Extensive work has been carried out using analytic, one-zone models of galaxy evolution \citep[e.g.,][]{Seok14, HM20}, which offer excellent physical insight but rely on over-simplified treatments of galaxy evolution. More recently, idealized zoom-in galaxy simulations have begun to incorporate PAHs \citep[][]{Narayanan23, curro26}, but despite capturing detailed grain physics, they cannot sample a cosmologically representative galaxy population. The most complete effort so far couples a dust-based PAH model with emission predictions in zoom-in hydrodynamic simulations \citep{Nara26}, but remains, by construction, limited to a modest number of individually simulated galaxies.

In this paper we present the first large-volume cosmological simulation of galaxy evolution\rev{, based on a semi-analytic model,} to combine a physically motivated, grain-size-resolved model of dust and PAH abundance with full radiative transfer (RT) post-processing to compute PAH emission across a wide redshift range, providing the statistical population needed to interpret the observed PAH scaling relations, and to extend these predictions to the regimes that current (JWST) and future (PRIMA) observations will probe. 
\rev{The paper is structured as follows. In Section \ref{sec:model} we describe our galaxy evolution model, including the treatment of the dust grain size distribution and PAHs, and the computation of their emission through RT with GRASIL. In Section \ref{sec:abundance} we present the predicted PAH abundances, their relation with metallicity and sSFR, the processes driving PAH build-up, and their cosmic evolution. In Section \ref{sec:emission} we analyze PAH emission, deriving a calibration to infer $q_{\rm PAH}$ from observables and presenting PAH scaling relations, which we compare with observations. In Section \ref{sec:overallpicture} we combine the main findings of the previous two sections into an overall picture of the joint evolution of PAH abundance and emission. We discuss our results in light of the existing literature, together with the main limitations of our model and possible improvements, in Section \ref{sec:discussion}, and summarize our conclusions in Section \ref{sec:conclusion}.}

\section{The model}\label{sec:model}
\label{sec:model}
\subsection{Galaxy formation and grain-size evolution}
\label{sec:galevo}

We adopt the public\footnote{\url{https://lgalaxiespublicrelease.github.io/index.html}} \textsc{L-Galaxies 2020} semi-analytic model \citep[SAM;][]{Henriques2020}, including the disc-instability and SMBH-growth modifications of \citet{2023MNRAS.521.6105P}, run on the dark matter merger trees of the \textsc{Millennium} simulation \citep[$L_{\rm box} = 500\,h^{-1}\,{\rm Mpc}$;][]{Millennium}. The model follows the standard set of astrophysical processes -- gas cooling, molecular-based star formation, stellar and AGN feedback, chemical enrichment from evolved stellar populations, merger-driven starbursts and disc instabilities -- described in full in the online documentation\footnote{\url{https://lgalaxiespublicrelease.github.io/Hen20_doc.pdf}}.

On top of this, we build on the \citet{Parente26} implementation of grain size evolution within the SAM, which self-consistently tracks the full dust grain size distribution (GSD) of both silicate and carbonaceous grains alongside the galaxy formation physics of the SAM. Dust grains are discretized into $N=32$ logarithmically spaced size bins per chemical species (carbonaceous and silicate, $a = 10^{-4}$--$10\,\mu\mathrm{m}$), and are seeded by AGB stars and core-collapse supernovae (SNe) with a lognormal size distribution peaked at large sizes ($a_{\rm peak}=0.1\,\mu\mathrm{m}$). Once injected into the ISM, the GSD of each galaxy evolves through four size-dependent processes. Two of these -- shattering and coagulation, arising from grain--grain collisions -- redistribute mass across grain sizes while conserving the total dust mass: shattering results from collisions between grains in the low-density diffuse medium, where relative velocities are high, while coagulation operates in dense giant molecular clouds (GMCs), where low relative velocities allow colliding grains to stick together. Within these dense clouds, grains also grow in mass through the accretion of gas-phase metals, a process that is more efficient \rev{per unit mass} for small grains\rev{, owing to their larger surface-to-volume ratio}. Finally, destruction by SN-driven shocks and thermal sputtering in hot gas erodes grains and returns metals to the gas phase. Together, these processes drive the GSD from a large-grain-dominated distribution at early times toward a flatter, MRN-like shape \citep{MRN} as galaxies build up their dust-to-metals ratio, consistent with the local dust mass function, the dust-to-gas--metallicity relation, and Milky Way extinction curves \citep{Parente26}.

In this framework, PAHs are identified with the smallest carbonaceous grains (below $15$ \AA, corresponding to $N_{\rm C}\approx 1000$ carbon atoms; \citealt{Draine21}) that reside in the diffuse phase of the ISM. This translates into a PAH mass fraction defined as \begin{equation} 
q_{\rm PAH} \equiv \frac{M_{\rm PAH}}{M_{\rm dust}} = \frac{M^{\rm C}_{\rm dust}(a \leq 15 \, \mathrm{\AA})\,\times (1-f_{\rm H_2})}{M_{\rm dust}}, \label{eq:qpah} \end{equation} 
which is therefore a simple function of the GSD and dust composition predicted by the simulation at any given time. The factor $(1-f_{\rm H_2})$, where $f_{\rm H_2}$ is the molecular fraction of the ISM gas, restricts PAHs to the diffuse medium, reflecting the assumption that aromatization occurs only in this phase, while aliphatization dominates in the dense phase \citep{Murga19}. Since both \rev{aromatization and aliphatization} timescales are typically short ($\lesssim 10^{6}\,{\rm yr}$) compared to the SAM evolutionary timesteps, the aromatic fraction can be taken as simply proportional to the diffuse gas fraction \citep{HM20}. The key point to highlight is that\rev{, by construction,} PAHs \rev{in our model} are not injected directly by stellar sources -- which instead produce large grains -- but are built up \emph{in situ}, primarily through the shattering of larger carbonaceous grains in the diffuse ISM, and subsequently grown or eroded by accretion, coagulation, and destruction. \rev{We stress that this is an assumption of our model rather than an observational constraint: aromatic features have been detected in the dusty outflows of evolved stars \citep[e.g.,][]{Lau22}.}

We acknowledge that this treatment of PAHs is likely an oversimplification, due both to the lack of some potentially important processes -- such as PAH formation in stellar ejecta or photoprocessing (see the discussion in Section \ref{sec:discussion}) -- and to our treatment of PAHs as just the small-size end of the carbonaceous GSD.
For instance, the use of spheres to model collisions -- as assumed in our shattering and coagulation framework -- may be inaccurate for PAHs of relatively few atoms. \citet{Hirashita23} tested this directly by decoupling PAH evolution from the ISM reprocessing applied to larger grains, and found that this actually improves the match to the Milky Way extinction curve and dust emission SED.
Nonetheless, we anticipate that our model predictions agree with most of the observed scaling relations, suggesting that this remains an adequate \textit{effective} description of PAH physics at the population level -- though this success does not automatically validate the underlying microphysics. Moving toward an explicitly molecular treatment (tracking distinct chemical species, ring structure, growth by specific chemical reactions) remains beyond the goal of our cosmological simulations, but is the natural direction for future refinement.

\subsection{SED modeling with GRASIL}\label{sec:grasil}

\label{sec:grasil}

In order to obtain a SED for our simulated galaxies we post-process the SAM outputs with the radiative transfer code GRASIL \citep{Silva98, Granato00}, post-processing the star formation history (SFH), metal enrichment history, and gas/dust content predicted by the SAM at any given output time. Stars are assumed to form inside optically thick GMCs and to progressively escape into the diffuse medium (cirrus) on a timescale $t_{\rm esc}=3\,{\rm Myr}$, so that young stellar populations are efficiently obscured while older populations are in the diffuse ISM. Radiative transfer inside the GMCs is solved explicitly
\citep{Granato94}, while an effective absorption optical depth is adopted for the diffuse component. We refer the reader to \citet{Parente25GV} for the full description of the coupling between the SAM and GRASIL.

The only substantial modification with respect to the \citet{Parente25GV} setup concerns the dust model itself. The SAM used here predicts a full GSD for each of the two chemical compositions (carbonaceous and silicate). This size distribution is discretized on a grid of
$N_{\rm bin}=16$ logarithmically-spaced size bins per chemical species and then passed to GRASIL.
For each of the 16 size bins per species, GRASIL adopts the optical properties (absorption and scattering efficiencies $Q_{\rm abs}(a,\lambda)$,
$Q_{\rm sca}(a,\lambda)$) of astronomical silicate and graphite grains from \cite{Draine84} and \cite{Laor93}. Given the relevance to this work, below we provide a more detailed summary of the treatment of PAHs in GRASIL, while referring to the original works for full details \citep{Silva98, Vega05}.

\subsubsection{Setting the PAH population}
\label{sec:pahgrasil}

PAH molecules are identified in GRASIL with the carbonaceous grain population with sizes $a < a_{\rm cut}=15\,\text{\AA}$ in the diffuse medium, and their abundance is read directly from the size-resolved carbonaceous dust mass predicted by the SAM.

As for PAHs, this radius is to be understood as an \emph{effective} radius, i.e. the radius of a sphere having the same volume as the grain, so that the number of carbon atoms in a carbonaceous grain is
$ N_{\rm C} \;=\; 460 \left( {a}/{10\,\text{\AA}} \right)^{3} \, $ having assumed a bulk density of graphitic carbon $\rho = 2.2\,{\rm g\,cm^{-3}}$ \citep{Draine21}. GRASIL, on the other hand, describes PAHs internally through their \emph{planar} radius $a_{\rm 2D}$, appropriate for the flat, catacondensed geometry of these molecules, for which $a_{\rm 2D} \simeq 0.9\,\sqrt{N_{\rm C}}\ \text{\AA}$ \citep{Vega05}. The two conventions are therefore related by $a_{\rm 2D} \simeq 19.3 \,(a/10\,\text{\AA})^{3/2}\,\text{\AA}$.

The shape of the PAH size distribution is imposed, and assumed to be a power law $\mathrm{d}n/\mathrm{d}a_{\rm 2D} \propto a_{\rm 2D}^{\,\beta_{\rm PAH}}$ with $\beta_{\rm PAH} = -3.5$\rev{, as in the original GRASIL implementation \citep{Silva98, Vega05},} between $N_{\rm C} = 20$ and the $N_{\rm C}$ corresponding to $a_{\rm cut}$.
In terms of the effective radius used elsewhere in this paper, the same distribution reads $\mathrm{d}n/\mathrm{d}a \propto a^{(3\beta_{\rm PAH}+1)/2} \propto a^{-4.75}$. Since the total PAH mass is set by the SAM, this assumption only affects how that mass is distributed among molecule sizes, and hence the relative strength of the individual mid-IR features -- a quantity we do not discuss, since our model makes no original prediction in this respect (see Section \ref{sec:discussion}).

The UV-to-optical absorption responsible for exciting the PAHs is computed adopting tabulated cross sections per carbon atom in the range $\lambda \simeq 0.05$--$0.5\,\mu$m following \cite{Silva98}. The mid-IR emission is modelled with the integrated cross sections of \citet{DraineLi07} for 24 bands in the range $3.3$--$18.92\,\mu$m, tabulated separately for neutral and ionized PAHs. Each band $i$ is characterised by its integrated strength per atom $\sigma_i$, central frequency $\nu_i$ and effective bandwidth $\Delta\nu_i$, and is reconstructed in the output spectrum as a Drude profile \citep{LiDraine01}.

The bands are further split into those due to C--C skeletal vibrations of the aromatic rings (6.2, 6.7, the 7.7\,$\mu$m complex and the features longward of $14\,\mu$m) and those due to C--H stretching or bending modes (3.3, 5.3, 5.7 and the 8.6, 11.3, 12.7, 13.5\,$\mu$m complexes) -- a distinction explicit in \citet{DraineLi07}. Since carbon atoms populate the whole area of a quasi-planar PAH disc while hydrogen atoms sit on its perimeter, the former scale with $N_{\rm C} \propto a_{\rm 2D}^{2}$ and the latter with $N_{\rm H} \propto a_{\rm 2D}$, with the hydrogen coverage of the available peripheral sites fixed to reproduce the C--H to C--C band ratio of the Galactic cirrus \citep{Vega05}. 

The relative weight of the neutral and ionized tabulations is set by the ionization fraction, parametrized as a function of PAH size alone \citep{Draine21, Hensley_astrodust}:
\begin{equation}
  f_{\rm ion}(a) \;=\; 1 - \frac{1}{1 + a/(10\,\text{\AA})} \, 
  \label{eq:fion}
\end{equation}
(with $a$ the effective radius), reflecting the fact that smaller PAHs have higher ionization potentials, and are therefore harder to ionize than larger ones. Throughout the following we denote by $\sigma_i(a) = f_{\rm ion}(a)\,\sigma_i^{\rm ion} + [1-f_{\rm ion}(a)]\,\sigma_i^{\rm neu}$ the resulting ionization-weighted cross section of band $i$. \rev{The PAH charge state is a key driver of the relative feature strengths \citep[e.g.,][]{DraineLi07}. Since our $f_{\rm ion}$ depends on size alone, we do not use our model to predict feature ratios, deferring a radiation-field-dependent treatment of PAH ionization to future work (Section \ref{sec:disc:photoproc}).}

Finally, we note that in GRASIL the PAH abundance is specified independently for the dense (GMC) and diffuse (cirrus) ISM phases. As has been standard usage of the code since \citet{Silva98}, we set the PAH abundance in the GMCs to a negligible value, so that the PAH emission of our galaxies originates entirely in the cirrus component -- consistent with the aromatization assumption made at the SAM level (Section \ref{sec:galevo}). 

\subsubsection{Stochastic PAH heating in the single-photon limit} \label{sec:pah_heating}

Because PAHs contain too few atoms to reach thermal equilibrium when embedded in the local radiation field \citep[e.g.,][]{Leger84, Draine85}, their emission cannot be computed with the equilibrium energy-balance approach used for the larger grain size bins. GRASIL treats PAH heating in the single-photon limit \citep[e.g.,][]{Leger89, Rigopoulou21, Richie25}: the average time between successive photon absorptions by a single PAH molecule is much longer than the time it takes the grain to radiatively cool back down, so the grain effectively absorbs one UV/optical photon at a time, is heated to a peak temperature, and then cools completely before the next photon arrives. Below we highlight the main steps of this single-photon heating and its implementation in GRASIL, which essentially follows \cite{Puget85} and \cite{XuDeZotti}.

(1) A PAH molecule gets energy from the local radiation field by absorbing photons of frequency $\nu$, which raises its temperature to $T_{\rm max}(a,\nu)$, set by the energy balance:
\begin{equation}
h\nu = \int_{T_{\rm min}=5\,{\rm K}}^{T_{\rm max}(a,\nu)} C(T,a)\,{\rm d}T \ ,
\label{eq:tmax_balance}
\end{equation}
with $C(T,a)$ the heat capacity of the PAH molecule from \citet{Leger87}.

(2) The grain then radiates this energy away through the mid-IR bands: the instantaneous power emitted in band $i$ at temperature $T$ is $4\pi B_{\nu_i}(T)\sigma_i(a) N_i(a)\Delta\nu_i$, where $\sigma_i(a)$ is the aforementioned ionization-weighted cross section, $N_i(a)$ is the number of atoms (C or H, depending on the feature) contributing to band $i$, and $\Delta\nu_i$ is its effective bandwidth. Summing over all the mid-IR features, the total instantaneous radiated power is obtained as $P_{\rm rad}(T,a) = \sum_i 4\pi B_{\nu_i}(T)\sigma_i(a) N_i(a)\Delta\nu_i$.

(3) Equating the rate of energy loss to $P_{\rm rad}(T,a)$ defines how the grain cools with time,
\begin{equation}
\frac{{\rm d}T}{{\rm d}t} = -\,\frac{P_{\rm rad}(T,a)}{C(T,a)} \ .
\label{eq:coolingtime}
\end{equation}

(4) Averaging over many such absorption-cooling cycles -- weighting the band-$j$ power fraction $w_j$ during a single cooling event by the actual rate at which photons of every energy are absorbed (set by the radiation field and the optical/UV cross sections) -- gives the time-averaged power radiated in band $j$ by a grain of size $a$,
\begin{equation}
\langle P_j(a)\rangle = \int {\rm d}\nu\;\dot N_{\rm abs}(a,\nu)
\int_{T_{\rm min}}^{T_{\rm max}(a,\nu)} w_j(T,a)\;C(T,a)\,{\rm d}T \ .
\label{eq:ellj}
\end{equation}
This guarantees by construction energy conservation between the power absorbed and the power re-emitted across the mid-IR bands.
The total band-$j$ luminosity is obtained by integrating $\langle P_j(a)\rangle$ over the PAH size distribution.
The individual bands are finally reconstructed as Drude profiles in wavelength for the output spectrum \citep{LiDraine01}.
\section{PAH abundance}\label{sec:pah_abundance}
\label{sec:abundance}

In this section we present our model predictions for the PAH abundance, focusing on its main scaling relations, the processes driving them, and its cosmic evolution.

\subsection{The PZR and PSR}
\label{sec:PZR}

\begin{figure}[]

    \centering
    \includegraphics[width=\columnwidth]{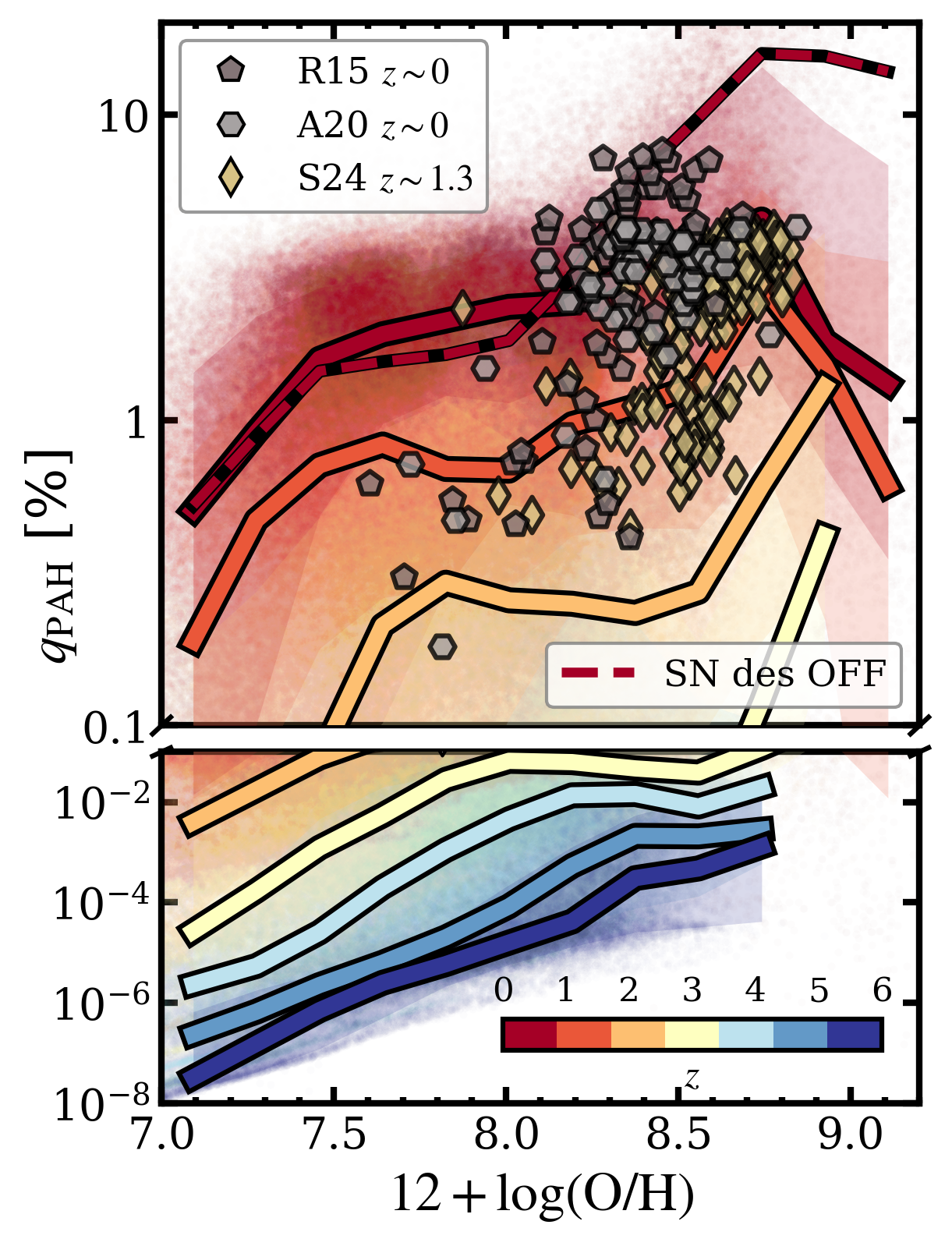}

    \caption{\textbf{The redshift evolution of the PZR.} 
    The figure shows the PZR ($q_{\rm PAH}$ vs metallicity) relation for our simulated galaxies across $0<z<6$. Alongside points, we report the median and $16-84$th percentile dispersion of the relation at each redshift as solid lines and shaded areas. The $z=0$ PZR of a run where grains destruction by SNe has been switched off is also shown as a dashed line. For reference we report also observational data from \cite{RR15, Aniano20, Shivaei24}.} 
    \label{fig:PZR:main}
\end{figure}


\begin{figure}[]

    \centering
    \includegraphics[width=\columnwidth]{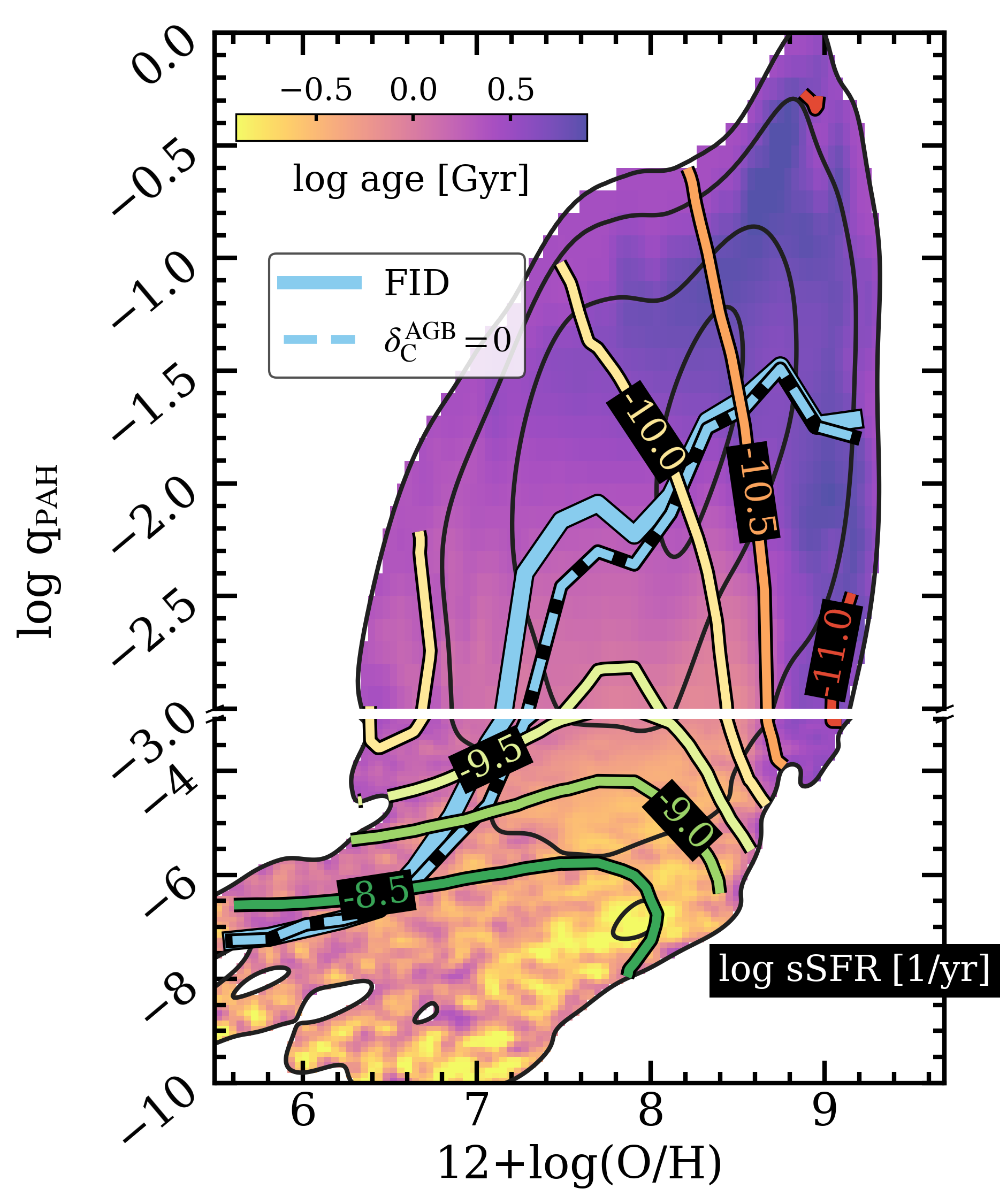}

    \caption{\textbf{Galaxy age is associated with the scatter of the PZR.} 
    The PZR for our simulated galaxies at all redshifts colored by the mass weighted stellar age. Black contours follow the distribution of galaxies in the diagram. Solid colored lines are medians of galaxies with similar sSFR, with the number on top being the median $\log {\rm sSFR}/{\rm yr^{-1}}$. The light cyan solid line represents the median of the full population, while the black-cyan dashed line is the median for galaxies where the production of carbon grains by AGB stars has been turned off.} 
    \label{fig:PZR:colorage}
\end{figure}


\begin{figure}[]

    
    \centering
    \includegraphics[width=\columnwidth]{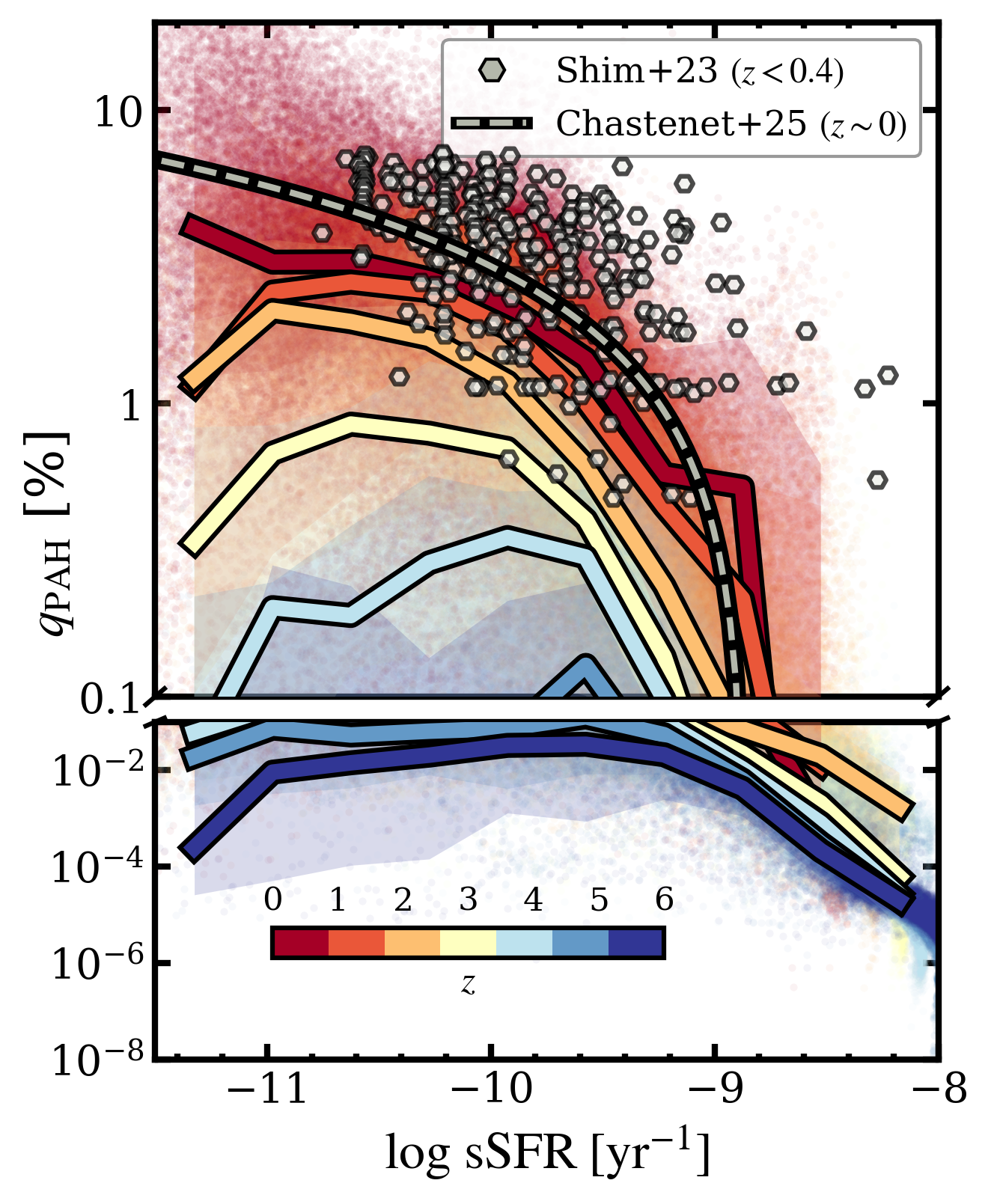}

    \caption{\textbf{The PAH mass fraction anti-correlates with the specific SFR.} Relation between $q_{\rm PAH}$ and specific star formation rate (sSFR) of our simulated galaxies at $0 < z < 6$, with medians and $16-84$th percentile dispersion represented as solid lines and shaded areas. We report observational results by \cite{Shim23} for $z < 0.4$ galaxies and the fit obtained by \cite{Chastenet25} from a sample of nearby galaxies.} 
    \label{fig:PZR:sSFR}
\end{figure}

We start by analyzing the abundance of PAHs and interpreting the relation between PAH abundance ($q_{\rm PAH}=M_{\rm PAH}/M_{\rm dust}$) and ISM metallicity, hereafter the PZR, shown in Figure \ref{fig:PZR:main}. Many observations have indeed found a positive correlation between $q_{\rm PAH}$ and $Z_{\rm ISM}$, and in particular a suppression of PAHs in low-metallicity galaxies \citep[e.g.,][]{Engelbracht05, Draine07, Shim23, Shivaei24, Whitcomb26}, although the physical origin of this trend has not yet been fully clarified.

Despite the large scatter, our model generally predicts an increase of $q_{\rm PAH}$ with the ISM metallicity, together with a higher normalization at lower redshift. At $z \lesssim 1$, the high-metallicity drop arises because in high SFR galaxies SN destruction  \citep[which is more effective on small grains; see][]{Parente26} overcomes the saturated metal accretion onto grains. 
\rev{Although SN rates are higher at cosmic noon, this only happens at low redshift because of the lower dense-gas fractions (see also Figure \ref{fig:PZR:experiments}) where accretion occurs, which at higher redshift can instead compensate for SN destruction. This is confirmed by Figure \ref{fig:PZR:main}, where switching off SN destruction at $z=0$ allows the PZR to rise above $12+\log(\rm O/H) \gtrsim 8.0$.}

The higher PAH abundance in high metallicity galaxies in our model can be understood in light of the ISM growth of grains, which dominates in metal enriched galaxies, building up their dust mass, which is eventually shattered to form small grains and PAHs -- which are just a fraction of small carbonaceous grains in our model (Equation \ref{eq:qpah}\rev{; see also Section \ref{sec:dust_proc}}). The greater abundance of small grains with increasing metallicity in our model has been already discussed in \cite{Parente26}, as well as both in analytic models based on conceptually similar dust size evolution models \citep[e.g.,][]{HM20}, and cosmological zoom-in simulations \citep{Nara26}.

We compare our model predictions with observational estimates from low-$z$ \textit{Spitzer}+\textit{Herschel} data \citep[][]{RR15, Aniano20} and $z \approx 1.3$ observations with JWST/MIRI multi-band photometry \citep{Shivaei24}. In all three cases, $q_{\rm PAH}$ is derived using the \cite{DraineLi07} dust models, fitted either to the mid- and far-IR photometry alone or, in the latter case, within the full panchromatic SED-fitting code \textsc{PROSPECTOR} \citep{PROSPECTOR}.

Our predictions are in good agreement with the observational determinations, although we note some low metallicity $z\sim 0$ observed galaxies suggesting a slightly steeper trend of the observed PZR compared to our local predictions, or equivalently, a higher \textit{critical metallicity} where the relation is expected to steepen. We however highlight that the $q_{\rm PAH}$ obtained from these commonly adopted methods in observations is not necessarily tracking the true physical abundance of PAHs, which is instead the output of our galaxy evolution simulation \citep[][C. Rubin et al., in prep.]{Nara26}.

While $q_{\rm PAH}$ correlates positively with metallicity, our model also predicts a strong dependence on galaxy age -- i.e., the mass-weighted age of the stellar population. Figure \ref{fig:PZR:colorage} shows this explicitly: at fixed metallicity, galaxies are more PAH-rich when older, or similarly, when their sSFR is lower. This is shown also in the $q_{\rm PAH}$--sSFR relation (hereafter the PSR) of Figure \ref{fig:PZR:sSFR}, where an anti-correlation emerges and steepens toward low redshift. The rising values of $q_{\rm PAH}$ in low-sSFR galaxies as redshift decreases simply reflects the same trend already noted above, namely that older, more metal-enriched galaxies are more PAH rich. This anti-correlation has already been reported observationally by \citet{Shim23}, from AKARI+\textit{Herschel} photometry of 373 low-$z$ star-forming galaxies, and by \citet{Chastenet25}, from resolved and integrated WISE+\textit{Herschel} SED fitting of $\sim 880$ nearby galaxies. The trend is well matched by our simulation, confirming that our model reproduces the strong sSFR dependence of PAH abundance found in these observations. \rev{At high sSFR ($\log {\rm sSFR/{\rm yr^{-1}}} \gtrsim -9.5$) our model underestimates some observed $q_{\rm PAH}$ by up to an order of magnitude. Such galaxies are young and dense-gas rich in our model, having had little time to shatter their grains down to PAH sizes. A PAH formation channel in dense gas (Section \ref{sec:disc:bottomup}) may be the origin of this discrepancy.}

Notably, both the PZR and the PSR emerge in our model without explicit PAH photodestruction and without direct PAH injection by stellar sources\rev{, although restricting PAHs to the diffuse phase (Section \ref{sec:galevo}) may partly mimic the former}. We discuss the implications of this result in the context of the alternative explanations proposed in the literature in Section \ref{sec:disc:alternatives}.

\subsection{Grain growth or shattering: what drives PAH abundance?}

\label{sec:dust_proc}


\begin{figure}[]

    \centering
    \includegraphics[width=0.99\columnwidth]{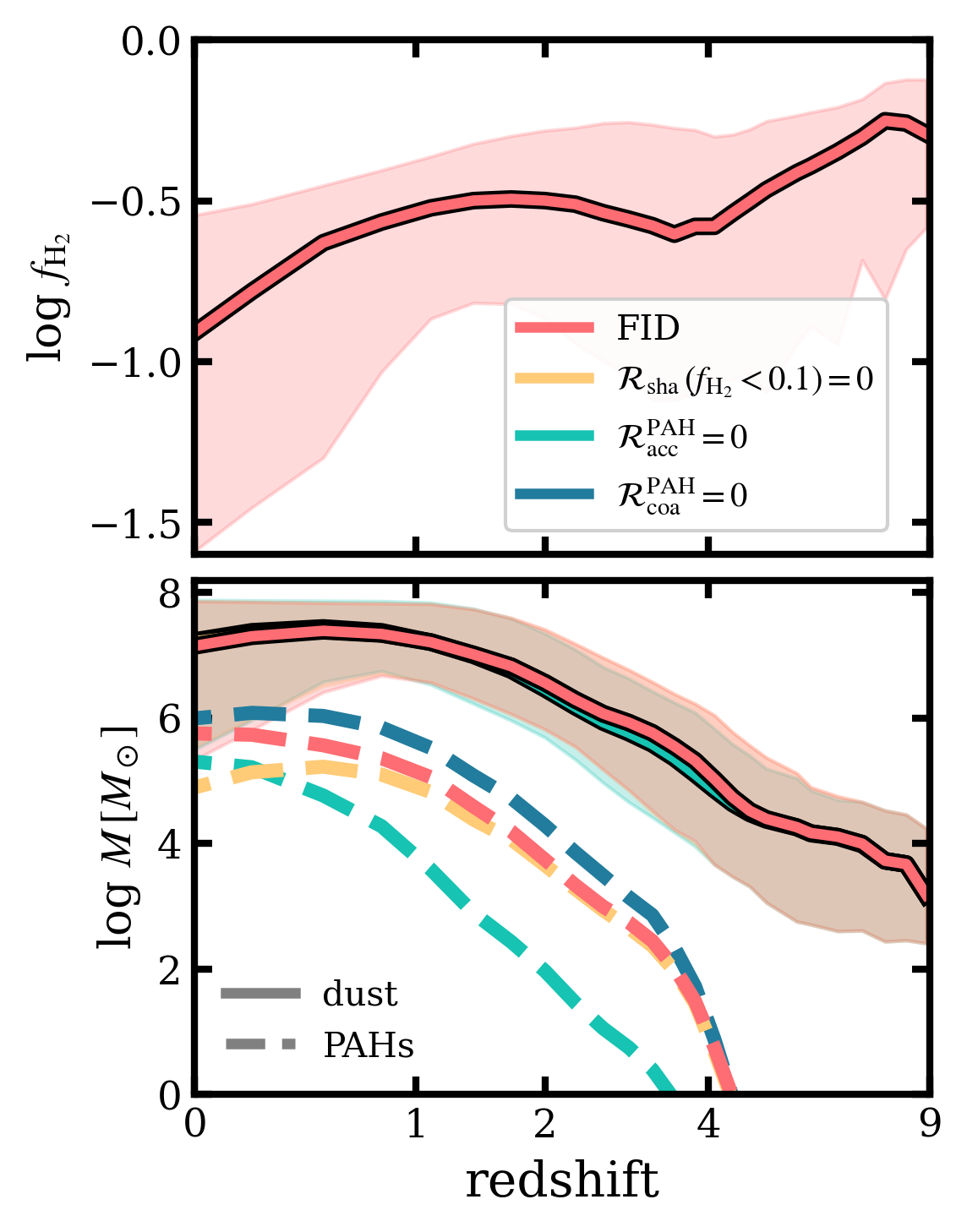}
        
    \caption{\textbf{\rev{The main channel of PAH mass growth shifts from accretion to shattering as galaxies lose their dense gas.}} Evolution of the dense ISM fraction $f_{\rm H_2}$ (top panel) and of dust and PAH mass (bottom panel) for a sample of simulated galaxies selected with $10^{10} < M_*^{z=0}/M_\odot < 8 \times 10^{10}$. Lines show median trends, with shaded regions (where present) indicating the $16$--$84$th percentile dispersion. In addition to the fiducial model (FID; red), we show \rev{three} runs: one in which shattering is shut off whenever $f_{\rm H_2}<0.1$ ($\mathcal{R}_{\rm sha}\,(f_{\rm H_2}<0.1)=0$), one in which accretion onto PAH-sized grains is turned off ($\mathcal{R}^{\rm PAH}_{\rm acc}=0$), \rev{and one in which coagulation of PAH-sized grains is turned off ($\mathcal{R}^{\rm PAH}_{\rm coa}=0$)}.}
    \label{fig:PZR:experiments}
\end{figure}


\begin{figure*}[]

    \centering
    \includegraphics[width=2\columnwidth]{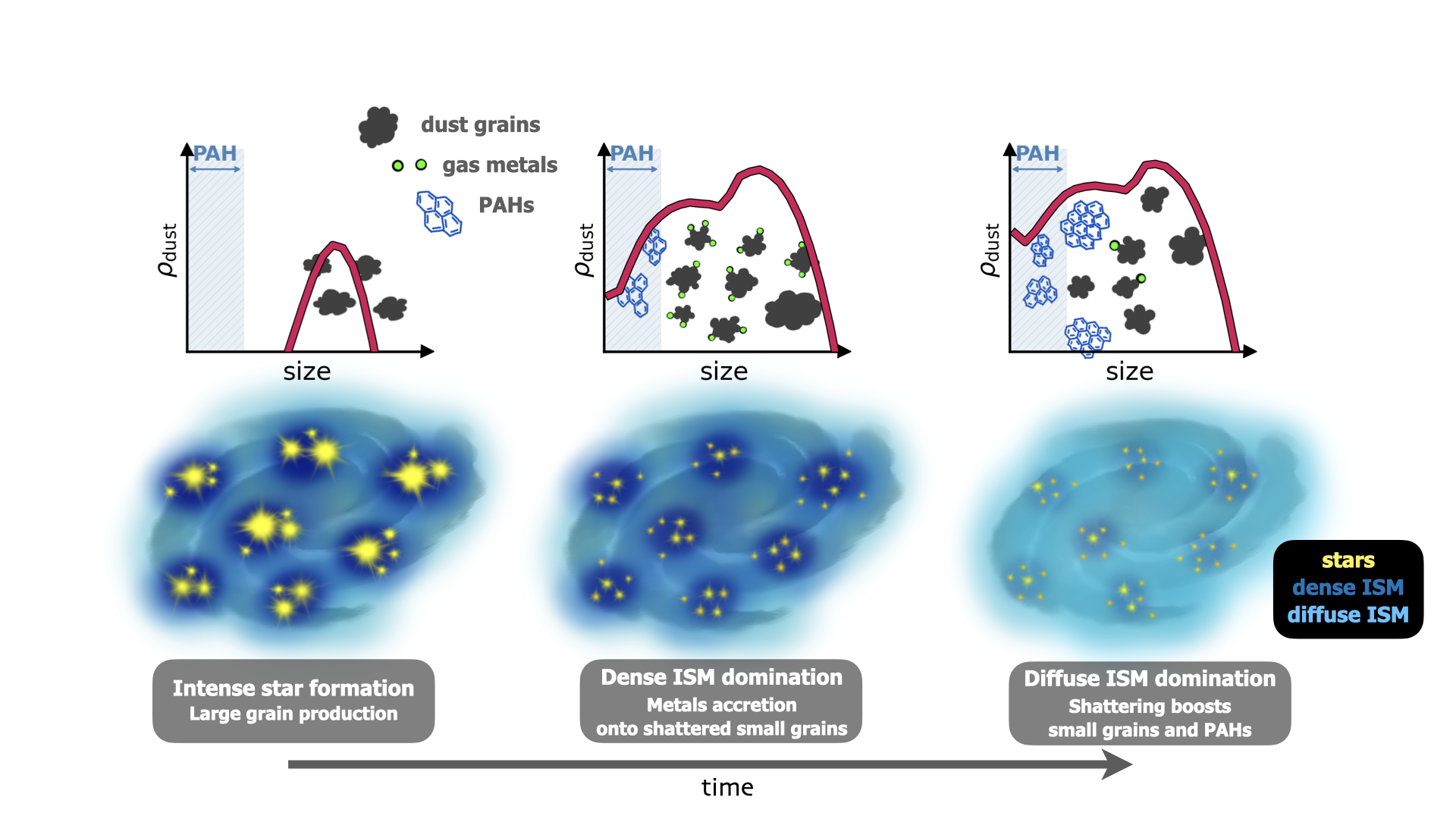}
    \caption{\textbf{Formation of PAHs by ISM shattering in a cartoon.} 
    The sketch outlines the life-cycle of grains (and PAHs) during different phases of galaxy evolution. The first star formation episodes contribute with typically large dust grains. While stars continuously replenish the ISM with large grains, the combined action of shattering and accretion boosts the total abundance of grains as well as the relative fraction of small grains. Far from the star formation dominated phase, when the galaxy is \textit{old}, there is a lack of molecular gas, so ISM grain growth is inhibited and shattering can further shift the distribution to smaller grains.} 
    \label{fig:cartoon}
\end{figure*}

\rev{In our model, the PZR and PSR emerge entirely from the ISM processing of the grain size distribution. The physical picture is as follows. Stars inject large grains, which are fragmented by grain--grain collisions in the diffuse ISM. The smallest carbonaceous fragments ($a < 15$\,\AA) constitute the PAH population (Equation \ref{eq:qpah}). In dense gas, these PAH-sized grains grow in mass by accreting gas-phase carbon, and can also coagulate to form larger grains. Both processes can move them above the $15$ \AA\ threshold, removing them from the PAH population. Accretion therefore does not form new PAHs from gas-phase precursors, as in bottom-up scenarios (Section \ref{sec:disc:bottomup}), but only adds mass to the fragments produced by shattering.}
This raises a natural question: what role \rev{do dust processes} play in shaping PAH abundances? Since these processes interact in a complex way, not easy to disentangle \citep{Parente26}, we run some simple numerical experiments to gain physical intuition about their individual contributions.

Figure \ref{fig:PZR:experiments} shows the redshift evolution of the dense ISM fraction ($f_{\rm H_2}$) and of dust and PAH masses for a sample of simulated galaxies. Alongside our fiducial model, we show \rev{three} other runs: $(i)$ shattering is shut off whenever $f_{\rm H_2}<0.1$ -- i.e., once the ISM becomes diffuse-phase dominated ($\mathcal{R}_{\rm sha}\,(f_{\rm H_2}<0.1)=0$); $(ii)$ metal accretion onto PAH-sized grains is turned off ($\mathcal{R}^{\rm PAH}_{\rm acc}=0$)\rev{; and $(iii)$ coagulation of PAH-sized carbonaceous grains is turned off ($\mathcal{R}^{\rm PAH}_{\rm coa}=0$)}.

These experiments identify two distinct regimes. \rev{Accretion onto PAH-sized grains} matters most at high redshift, while switching it off reduces the $z \approx 0$ PAH abundance by only a factor of $\approx 3$. On the other hand, shattering in the diffuse phase only becomes \rev{the main driver of PAH mass growth} at $z \lesssim 1$, once galaxies become dense-gas-poor (top panel). There, shattering shifts the grain size distribution toward smaller sizes, so the PAH abundance grows even as the total dust mass declines slightly. \rev{Although accretion can also move PAH-sized grains above 15\,\AA, its net effect on the PAH abundance is positive. Coagulation instead acts as a minor sink at all redshifts: switching it off increases the PAH abundance by less than a factor of $\approx 2$, without changing its evolution.}

Taken together -- and treating this as a qualitative exercise -- \rev{shattering always provides the PAH seeds, since in our framework stars only produce large grains, while the process that mainly builds up PAH mass changes with time: at high redshift, when the dense-gas fraction is highest, it is the growth of these seeds by accretion, while at low redshift, as $f_{\rm H_2}$ declines, it is continued shattering of larger grains.}

\rev{In our model, PAH formation therefore follows a top-down route, sketched in Figure \ref{fig:cartoon}.} Early star formation enriches the ISM with large grains, subsequent accretion \rev{increases} the dust abundance, and shattering then progressively \rev{shifts} the size distribution toward smaller grains. At late times, the ISM of galaxies becomes increasingly diffuse-phase dominated, which favors the high-velocity collisions needed to fragment grains and produce PAHs. \rev{A similar picture emerges in} \citet{Nara26} using a conceptually similar dust evolution model embedded in higher resolution hydrodynamic simulations, which capture ISM physics -- and shattering in particular -- in a better way (Section \ref{sec:disc:galevo}).

\rev{This picture also explains the anti-correlation between $q_{\rm PAH}$ and sSFR (Figure \ref{fig:PZR:sSFR}), which arises from two effects in our model. First, galaxies with lower sSFR are typically older, and have therefore had more time to grow and shatter their grains down to PAH sizes. Second, since PAHs are restricted to the diffuse phase (Equation \ref{eq:qpah}), galaxies with high sSFR, which are rich in dense gas, host a smaller fraction of their PAH-sized grains in the diffuse phase.}

\subsection{The redshift evolution of PAH abundance}
\begin{figure}[]

    \centering
    \includegraphics[width=0.99\columnwidth]{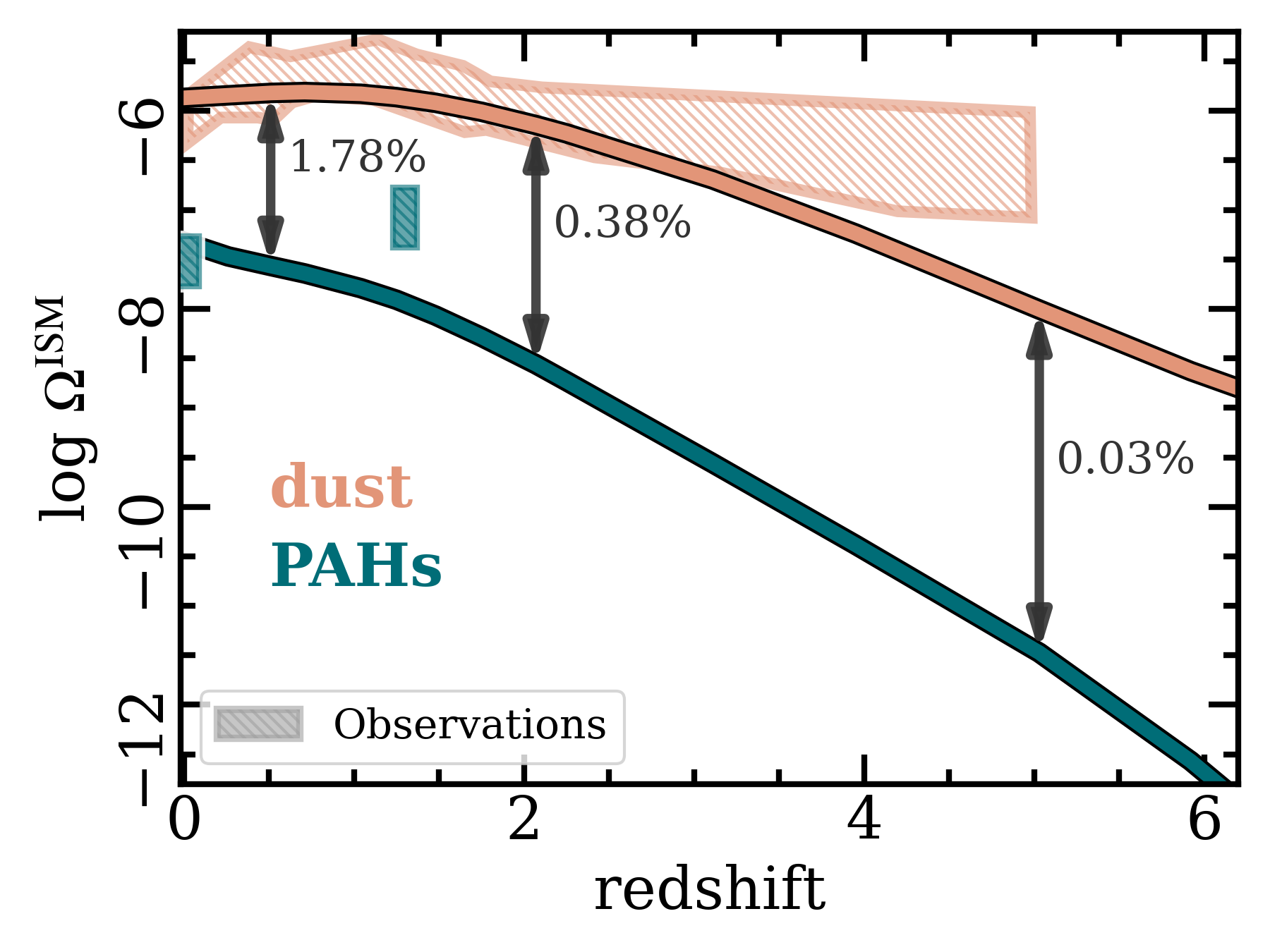}

    \caption{\textbf{Dust and PAHs evolve differently across cosmic time, with PAHs growth being steeper.} Redshift evolution of the cosmic ISM abundance parameter, $\Omega^{\rm ISM}$, for dust and PAHs. Solid lines show model predictions, while hatched regions show observationally based estimates. The $\Omega^{\rm ISM}_{\rm dust}$ compilation is taken from \cite{Parente25rev}. The $\Omega^{\rm ISM}_{\rm PAH}$ values are obtained by rescaling the dust abundance with $q_{\rm PAH}$ values at $z=0$ and $z\approx1.3$ from \cite{Aniano20} and \cite{Shivaei24}, respectively, convolved with the mass-metallicity relation, star forming main sequence, and stellar mass function (see text for details). Arrows and percentages indicate the average $q_{\rm PAH}$ predicted by the model at various redshifts.} 
    \label{fig:omegaPAH}
\end{figure}

To conclude this section, we present our predicted redshift evolution of the total PAH abundance\footnote{The parameter commonly used to indicate cosmic abundances is $\Omega=\rho/\rho_{\rm c,0}$, where $\rho$ is the comoving mass density (in our case, for PAHs or dust), and $\rho_{\rm c,0}=2.775\,h^2\times10^{11}\,M_\odot/{\rm Mpc}^3$ is the critical density of the Universe today.}, $\Omega^{\rm ISM}_{\rm PAH}$, alongside the dust abundance $\Omega^{\rm ISM}_{\rm dust}$, both shown in Figure \ref{fig:omegaPAH}. This comparison highlights a difference already noted in the previous sections: the PAH abundance rises more steeply than dust abundance with cosmic time. This steeper evolution reflects two combined effects: the assumption that PAHs reside exclusively in the non-dense ISM phase, whose mass fraction increases towards lower redshift, and the progressive reprocessing of the dust size distribution towards smaller grains as shattering becomes increasingly efficient.

Observationally motivated estimates of $\Omega^{\rm ISM}_{\rm dust}$ already exist \citep{Peroux20rev}, often obtained by integrating the dust mass function \citep[DMF; e.g.,][]{Driver2018, Pozzi2020, Traina24, Berta25}. An analogous estimate of $\Omega^{\rm ISM}_{\rm PAH}$ is still missing, mainly because too few galaxies beyond the local Universe have PAH detections to construct a PAH mass function. Here we take an admittedly simplified approach to translate PZR observations into a first estimate of $\Omega^{\rm ISM}_{\rm PAH}$.

We estimate a cosmic-averaged PAH mass fraction, $\langle q_{\rm PAH}\rangle(z)$, at $z=0$ and $z\approx1.3$ by convolving the observed PZR \citep[][taking the step functions reported in the latter work]{Aniano20, Shivaei24} with the star-forming stellar mass function \citep[$\phi(M_*,z)$, from][]{Muzzin13, Moffett16}. We use the mass--metallicity relation $Z(M_*,\mathrm{SFR},z)$ from \cite{Mannucci10} together with the star-forming main sequence $\mathrm{SFR}(M_*,z)$ from \cite{Speagle14} to map each stellar mass onto a characteristic gas-phase metallicity, and hence onto a corresponding $q_{\rm PAH}$. The stellar-mass-weighted\footnote{\rev{Since PAHs are stochastically heated, observed $q_{\rm PAH}$ values are likely effectively SFR-weighted. Here we adopt a mass weighting, more appropriate for a physical estimate of the cosmic PAH abundance.}} PAH mass fraction at redshift $z$ is then
\begin{equation}
\langle q_{\rm PAH}\rangle_{M_*}(z) =
\frac{\displaystyle\int
q_{\rm PAH}\big(Z(M_*,z)\big)\, M_*\, \phi(M_*,z)\, \mathrm{d}\log M_*}
{\displaystyle\int
M_*\, \phi(M_*,z)\, \mathrm{d}\log M_*},
\label{eq:qpah_mass_weighted}
\end{equation}
integrated over $8 \le \log(M_*/M_\odot) \le 12$. Using the values reported in the aforementioned works, our approach yields $\langle q_{\rm PAH}\rangle = 3.6\%$ and $3.2\%$ at $z=0$ and $z=1.3$, respectively. We obtain $\Omega^{\rm ISM}_{\rm PAH}$ by simply rescaling the observed $\Omega^{\rm ISM}_{\rm dust}$ compilation \citep{Parente25rev} by these fractions\footnote{\rev{The resulting uncertainty on  $\Omega^{\rm ISM}_{\rm PAH}$ therefore reflects only the spread in $\Omega^{\rm ISM}_{\rm dust}$, and does not include uncertainties of the adopted PZR, mass–metallicity relation, and stellar mass function, making our estimate a lower limit on the true uncertainty.}}.

The resulting estimate agrees well with our model in the local Universe, but disagrees by almost an order of magnitude at $z\approx1.3$, where the model falls below the observational value. This tension likely arises from the assumed shape of the PZR: the step function adopted here has a constant floor of $q_{\rm PAH}\approx1\%$ at low metallicity \citep{Shivaei24}, whereas in our model $q_{\rm PAH}$ can drop down to $0.1\%$ (Figure \ref{fig:PZR:main}). In addition, the observed PZR is necessarily biased towards PAH-detected galaxies, likely excluding objects with very low $q_{\rm PAH}$.
\rev{A PAH formation channel operating in dense gas, absent from our model (Section \ref{sec:disc:bottomup}), could also contribute to this tension, since it would raise $q_{\rm PAH}$ in the dense-gas rich galaxies that dominate at $z \gtrsim 1$.}
A more robust determination of $q_{\rm PAH}$ across a larger galaxy sample -- likely achievable with PRIMA \citep[][]{Yoon25} -- is therefore needed to test the predicted evolution of $\Omega^{\rm ISM}_{\rm PAH}$ and any difference in shape relative to $\Omega^{\rm ISM}_{\rm dust}$.


\section{PAH emission}
\label{sec:emission}
We now move to analyze emission properties of PAHs in our simulated galaxies.

\subsection{PAH Feature Extraction with PAHFIT}
\label{sec:pahfit}

\begin{figure}[]

    \centering
    \includegraphics[width=\columnwidth]{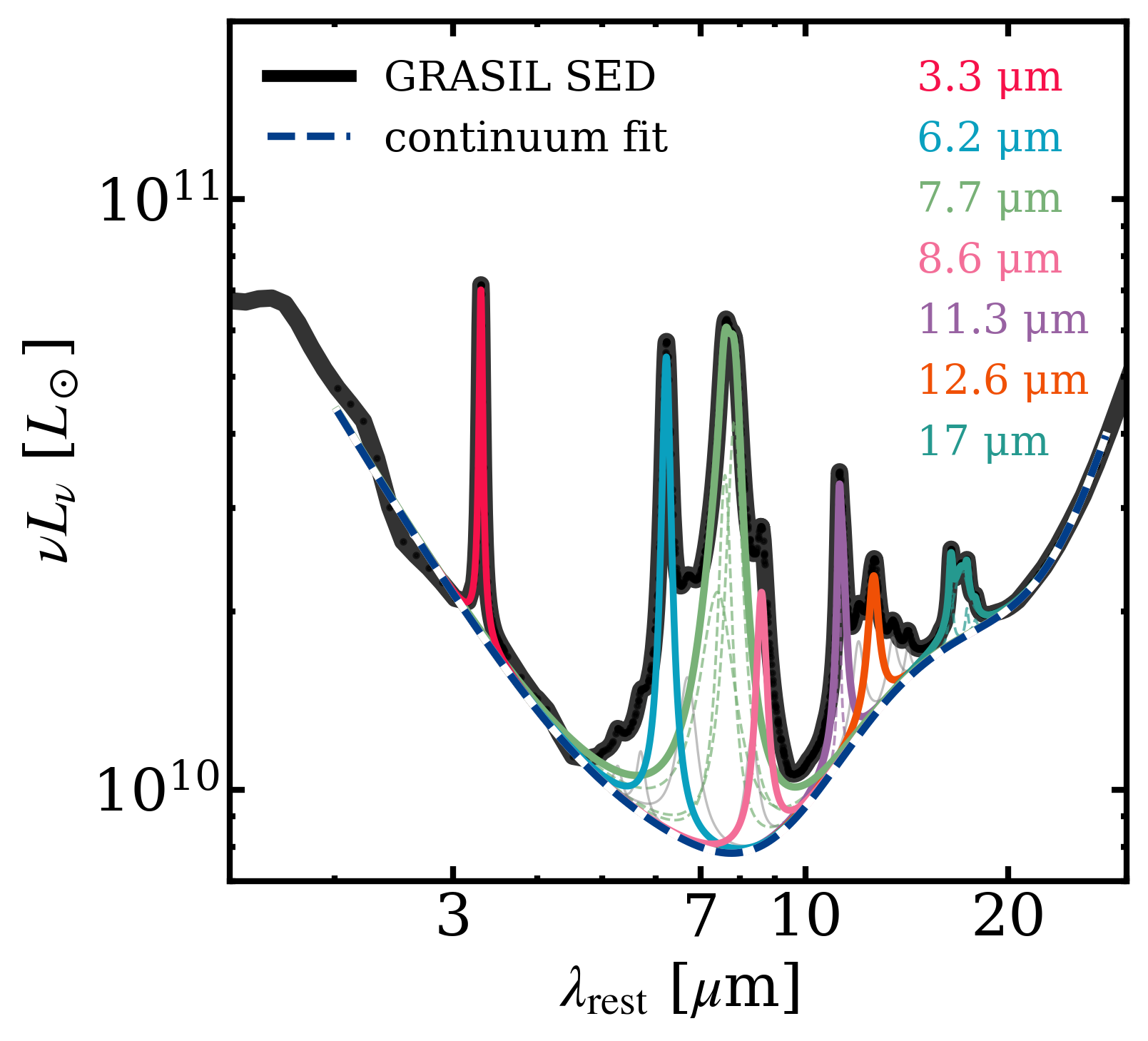}

    \caption{\textbf{PAH features extraction with PAHFIT.} 
    PAHFIT decomposition of a representative GRASIL SED (solid black line) for a galaxy at $z=0.0$. The fitted continuum is shown as a dashed blue line, and some of the main extracted PAH features are shown with different colors, with the individual sub-components of blended complexes plotted as dashed lines of the same color.} 
    \label{fig:PAHfit}
\end{figure}
To extract PAH feature strengths from our simulated mid-IR spectra, we decompose the GRASIL SEDs using \textsc{PAHFIT}\footnote{\url{https://pahfit.readthedocs.io/en/latest}} \citep{Smith07}, a spectral decomposition tool designed to separate the blended continuum, line, and dust-feature components of mid-IR galaxy spectra. Like other tools commonly adopted in the observational literature (e.g., CAFE; \citealt{CAFE}), PAHFIT allows a direct and fair comparison with observed galaxy spectra, and it returns individual fluxes for each PAH complex rather than a single blended luminosity.

PAHFIT models the observed mid-IR spectrum as the sum of a stellar continuum ($T=500$--$5000\,$K blackbody components), a set of modified-blackbody dust continuum components spanning $35$--$300\,$K, unresolved atomic and molecular emission lines, and the PAH emission features themselves, each represented as a Drude profile, $I_\nu(\lambda) = b\,\gamma^2 / \left[(\lambda/\lambda_0 - \lambda_0/\lambda)^2 + \gamma^2\right]$, where $\lambda_0$ is the central wavelength of the feature, $\gamma$ is its fractional width, and $b$ is the peak intensity. 
We hold the central wavelengths and widths of all Drude components fixed at their default PAHFIT values, allowing only the amplitudes to vary during the fit.

PAHFIT fits the major PAH features at $3.3$, $6.2$, $7.7$, $8.6$, $11.3$, $12.7$, and $17\,\mu$m, together with a number of weaker features. Several of the major complexes are themselves blends of multiple sub-features -- the $7.7\,\mu$m complex, for instance, is decomposed into three Drude components centered at $7.42$, $7.60$, and $7.85\,\mu$m -- and we sum the sub-component powers belonging to a given complex to obtain the total luminosity. We define the total PAH luminosity $L_{\rm PAH}$ as the sum of the integrated powers of all Drude features returned by the fit. In Figure \ref{fig:PAHfit} we show a representative PAHFIT decomposition of the mid-IR SED of one of our simulated galaxies.

\subsubsection{Selection criterion}

\rev{In applying the fitting procedure across our full galaxy sample, we find that the fit is not always physically meaningful. When the PAHFIT continuum components (i.e., its set of modified blackbodies) cannot fully reproduce the shape of the GRASIL continuum, the fit can compensate by using broad PAH features to model part of the continuum.} Such spurious detections must be excluded before any feature luminosity is inferred. For this reason, we retain only features whose peak luminosity substantially exceeds the local continuum (by $\gtrsim 50\%$). 

Moreover, we restrict our sample to galaxies with clear detections of both the $3.3\,\mu$m and $7.7\,\mu$m features. Besides being among the most luminous and best detected PAH bands, we find that, particularly at high redshift, warm dust \rev{heated by star formation\footnote{We note that AGN emission is not included in our model SEDs.}} can become sufficiently luminous in the rest-frame mid-IR to outshine the $7.7\,\mu$m complex entirely. This is itself a noteworthy physical prediction of our RT modeling, and one that may carry implications for feature-resolved observational studies of high-redshift galaxies.

Because a full exploration of this effect goes beyond the scope of this work, we restrict our analysis to galaxies in which both features \rev{satisfy our adopted criterion}. We defer a detailed investigation of the warm-dust/PAH interplay\rev{, and of the galaxies where this is relevant,} to future work\rev{, for which PRIMA will be crucial by probing the rest-frame mid-IR of high-redshift galaxies}. Here we only caution that the results presented below\rev{, particularly at $z \gtrsim 3$,} are conditioned on this selection, and should be interpreted as representative only of galaxies with clear PAH emission in both bands. 
\rev{Excluded galaxies are mainly gas-poor, low-SFR systems at $z \lesssim 1$, and young, high-sSFR galaxies at $z \gtrsim 2$, where the PAH features are often outshone by warm dust.}

\rev{We note that our selection criteria are designed to ensure a reliable feature extraction from our model spectra, rather than to mimic an observational selection. A more observationally motivated comparison will be the subject of future work.}

\subsection{From $L_{\rm PAH}/L_{\rm IR}$ to $q_{\rm PAH}$}


\begin{figure}[]

    \centering
    \includegraphics[width=0.99\columnwidth]{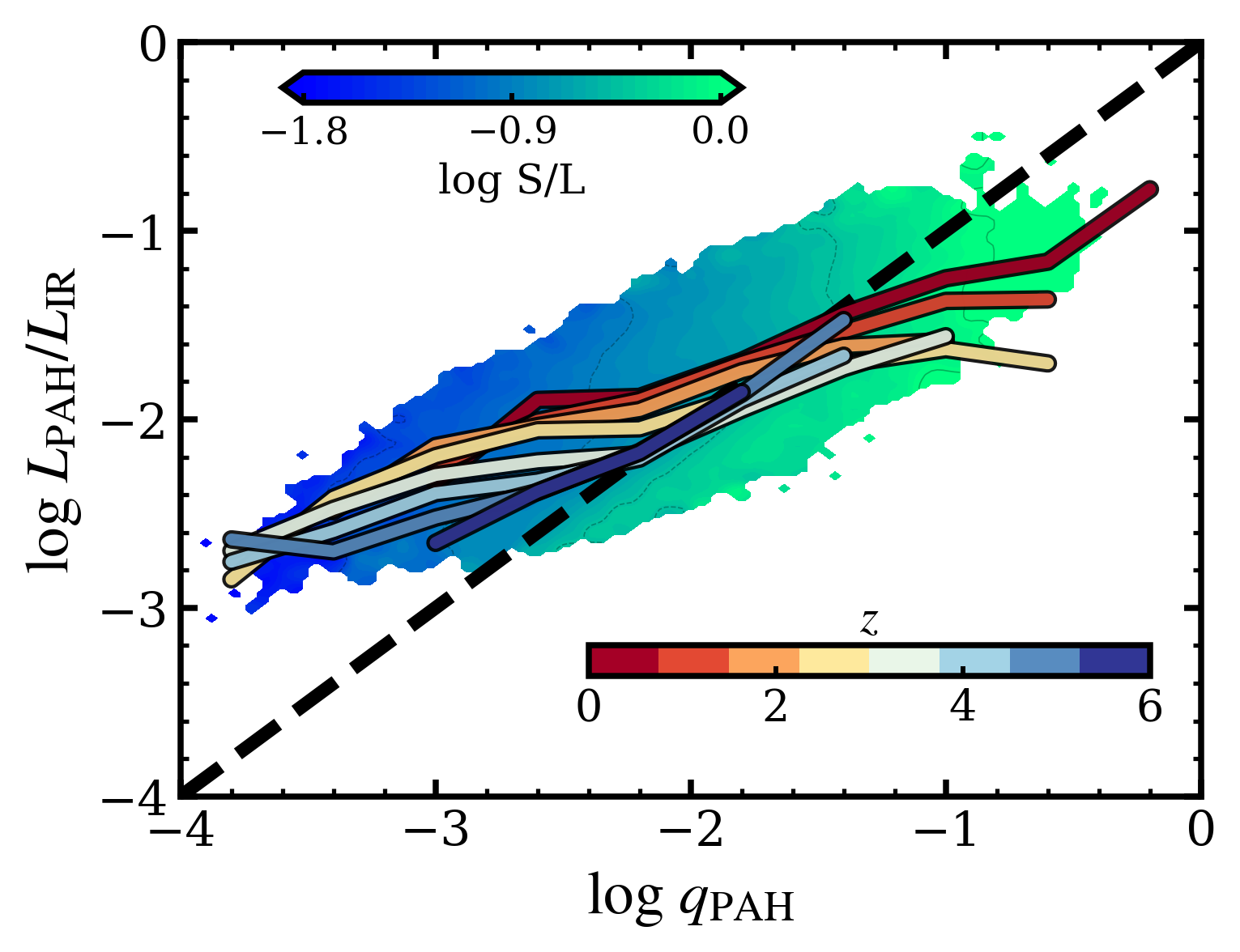}

    \caption{\textbf{PAH luminosity fraction alone is a noisy proxy for PAH mass fraction.} Relation between the PAH luminosity fraction $L_{\rm PAH}/L_{\rm IR}$ and PAH mass fraction $q_{\rm PAH}$. Solid lines represent medians at different redshift, while each galaxy is color coded according to their small-to-large dust mass ratio.} 
    \label{fig:PAHlum_mass_frac}
\end{figure}

We begin by examining how well the PAH luminosity fraction\footnote{Here the IR luminosity is defined as the integral of the SED between $8-1000\,\mu{\rm m}$.}, $L_{\rm PAH}/L_{\rm IR}$, traces the PAH mass fraction $q_{\rm PAH}$ introduced in Section \ref{sec:galevo}. Figure \ref{fig:PAHlum_mass_frac} shows this relation for our simulated galaxies at various redshifts. The two quantities are well correlated, although at fixed $q_{\rm PAH}$ the luminosity fraction can scatter by up to $\gtrsim1\,{\rm dex}$. \rev{We note the little redshift evolution of the relation, since the radiation field intensity, which sets both $L_{\rm PAH}$ and $L_{\rm IR}$, mostly cancels out in their ratio (unlike in the $L_{\rm PAH}-M_{\rm PAH}$ relation; see Section \ref{sec:PAHscaling}).}

The scatter of the relation is driven by competition to absorb the same UV photons between PAHs and other small grains, since both populations absorb efficiently at similar wavelengths. At fixed $q_{\rm PAH}$, galaxies with a higher small-to-large grain mass ratio (${\rm S/L}$, with small grains defined as $a<0.015 \, \mu{\rm m}$) show a lower $L_{\rm PAH}/L_{\rm IR}$. An abundant small-grain population absorbs UV photons that would otherwise be absorbed by PAHs, reducing the energy reprocessed into the PAH features. Conversely, when the overall grain population is dominated by large grains -- which absorb less efficiently per unit mass at UV energies -- a larger fraction of the UV radiation remains available to the PAHs, and $L_{\rm PAH}/L_{\rm IR}$ increases accordingly. \rev{Since S/L generally increases with $q_{\rm PAH}$, this also makes the relation shallower than the 1:1 relation.} We demonstrate this explicitly with dedicated numerical RT experiments in Appendix \ref{app:SEDexp}.

Finally, we note that $q_{\rm PAH}$ and the small-to-large mass ratio, while closely related (modulo the diffuse-phase factor; see Section \ref{sec:galevo}), are not necessarily tightly correlated. In our shattering-driven top-down formation scenario of small grains and PAHs, the S/L ratio (threshold at $a=0.015 \, \mu{\rm m}$) grows earlier when compared to the PAH population (threshold at $a=0.0015 \, \mu{\rm m}$). Galaxies at fixed $q_{\rm PAH}$ can therefore host different amounts of this competing small-grain population, contributing to the scatter in $L_{\rm PAH}/L_{\rm IR}$ discussed above.


\begin{figure}[]

    \centering
    \includegraphics[width=0.99\columnwidth]{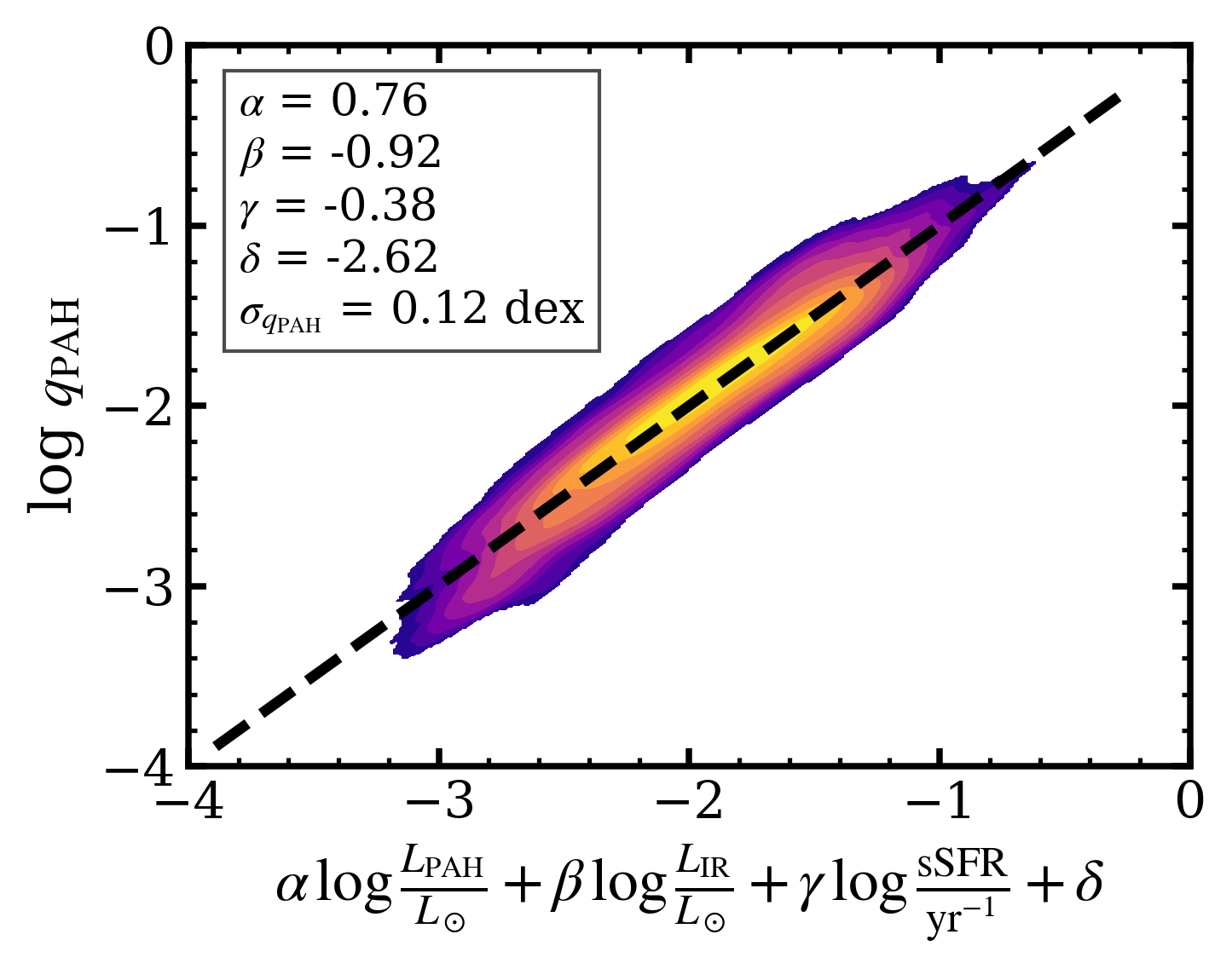}

    \caption{\textbf{The PAH mass fraction is well recovered from a four parameters fit.} Performance of a linear combination of PAH luminosity, IR luminosity, and specific SFR in recovering the PAH mass fraction $q_{\rm PAH}$. The best-fit coefficients and the typical scatter of the relation are reported.} 
    \label{fig:fit}
\end{figure}

Although the scatter in the $L_{\rm PAH}/L_{\rm IR}$--$q_{\rm PAH}$ relation discussed above is not negligible, the relation is sufficiently tight that, informed by our simulations, we can exploit the observable luminosity to directly infer the PAH mass fraction. Motivated by the role of the age of the galaxy in setting the PAH abundance (Section \ref{sec:PZR}), we additionally include the sSFR in the fit.\\
We fit a linear relation in log space of the form
\begin{equation}
    \log q_{\rm PAH} = \alpha \, \log \frac{L_{\rm PAH}}{L_\odot}
    + \beta \, \log \frac{L_{\rm IR}}{L_\odot}
    + \gamma \, \log \frac{\rm sSFR}{\rm yr^{-1}} + \delta,
    \label{eq:qpah_fit}
\end{equation}
with coefficients $\alpha$, $\beta$, $\gamma$, and $\delta$ determined via ordinary least squares in log space. The best-fit values, obtained from simulated galaxies spanning $z = 0-6$, are $\alpha = 0.76$, $\beta = -0.92$, $\gamma = -0.38$, and $\delta = -2.62$. The performance of the fit is shown in Figure \ref{fig:fit}.

We quantify the scatter of the relation using the median absolute deviation of the residuals ($\Delta \equiv \log q_{\rm PAH,\,true} - \log q_{\rm PAH,\,pred}$), scaled by a factor of $1.48$ to an estimator, $\sigma_{q_{\rm PAH}}$, equivalent to the standard deviation of a Gaussian distribution. We find $\sigma_{q_{\rm PAH}} = 0.12$ dex, meaning $q_{\rm PAH}$ can typically be recovered from $L_{\rm PAH}$, $L_{\rm IR}$, and sSFR to within a factor of $\approx 1.3$. \rev{We note that Equation \ref{eq:qpah_fit} does not include an explicit redshift dependence, in order to keep the fit dependent only on physical galaxy properties. Redshift evolution is nonetheless implicitly captured, e.g., through the sSFR.}

The relation reported here uses the total PAH luminosity, summed over all individual features. We do not provide feature specific relations, since we do not offer original predictions on the relative importance of individual features (Section \ref{sec:pahgrasil}). Observers wishing to apply Equation \ref{eq:qpah_fit} using a single PAH feature (e.g., the widely used $3.3$, $7.7$, or $11.3\,\mu$m features) could still do so by adopting empirically or theoretically motivated total-to-feature luminosity ratios \citep[e.g.,][]{Smith07, DraineLi07, Shipley16}.  Further analysis of the relationship between the luminosities of PAHs and mass fractions will be presented by C. Rubin et al. (in prep.).

\subsection{PAH scaling relations}

\label{sec:PAHscaling}

\begin{figure*}[t]

    \centering

    \begin{subfigure}{0.45\textwidth}
        \centering
        \includegraphics[width=\linewidth]{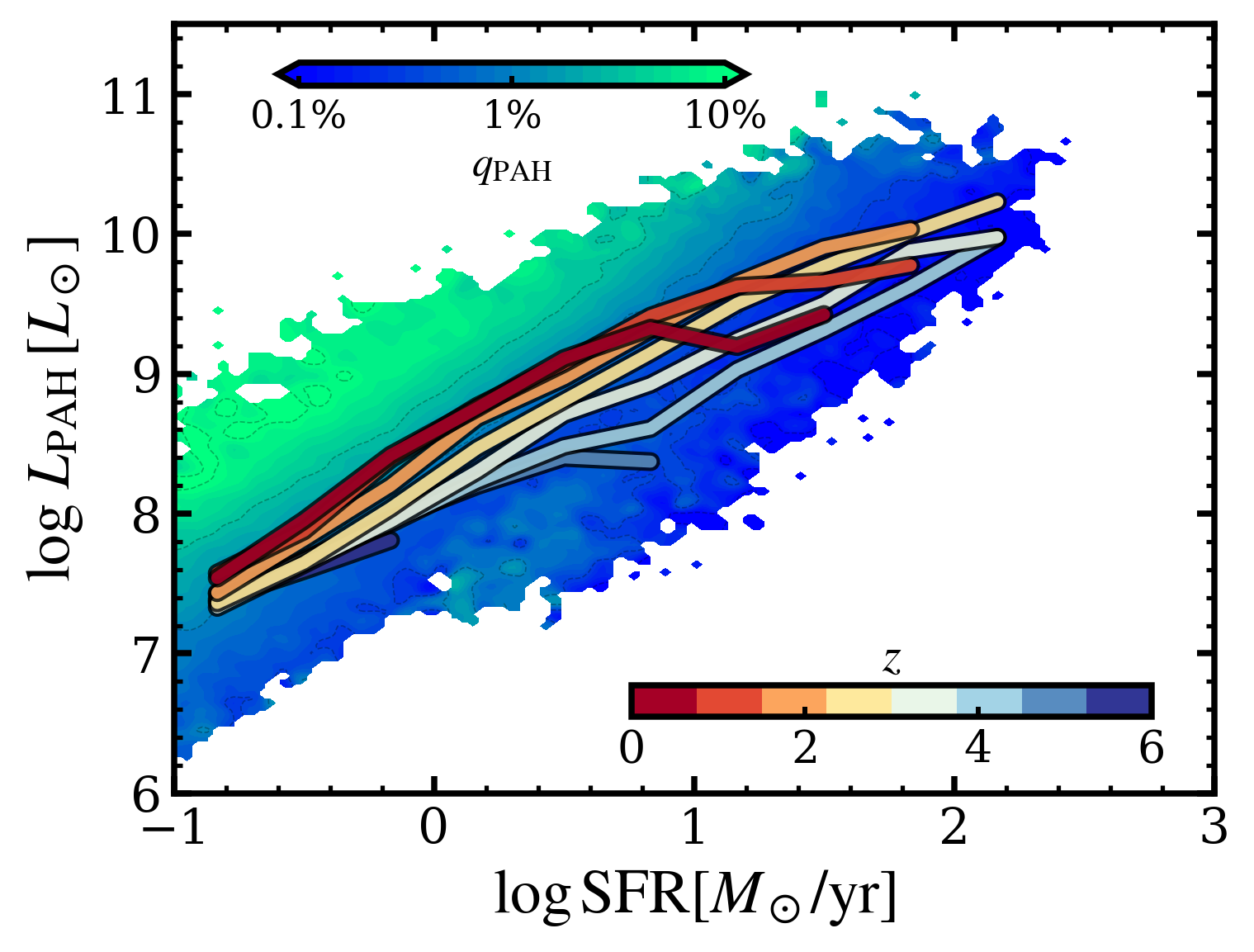}
        \caption{$L_{\rm PAH}$ vs SFR}
    \end{subfigure}
    \hfill
    \begin{subfigure}{0.45\textwidth}
        \centering
        \includegraphics[width=\linewidth]{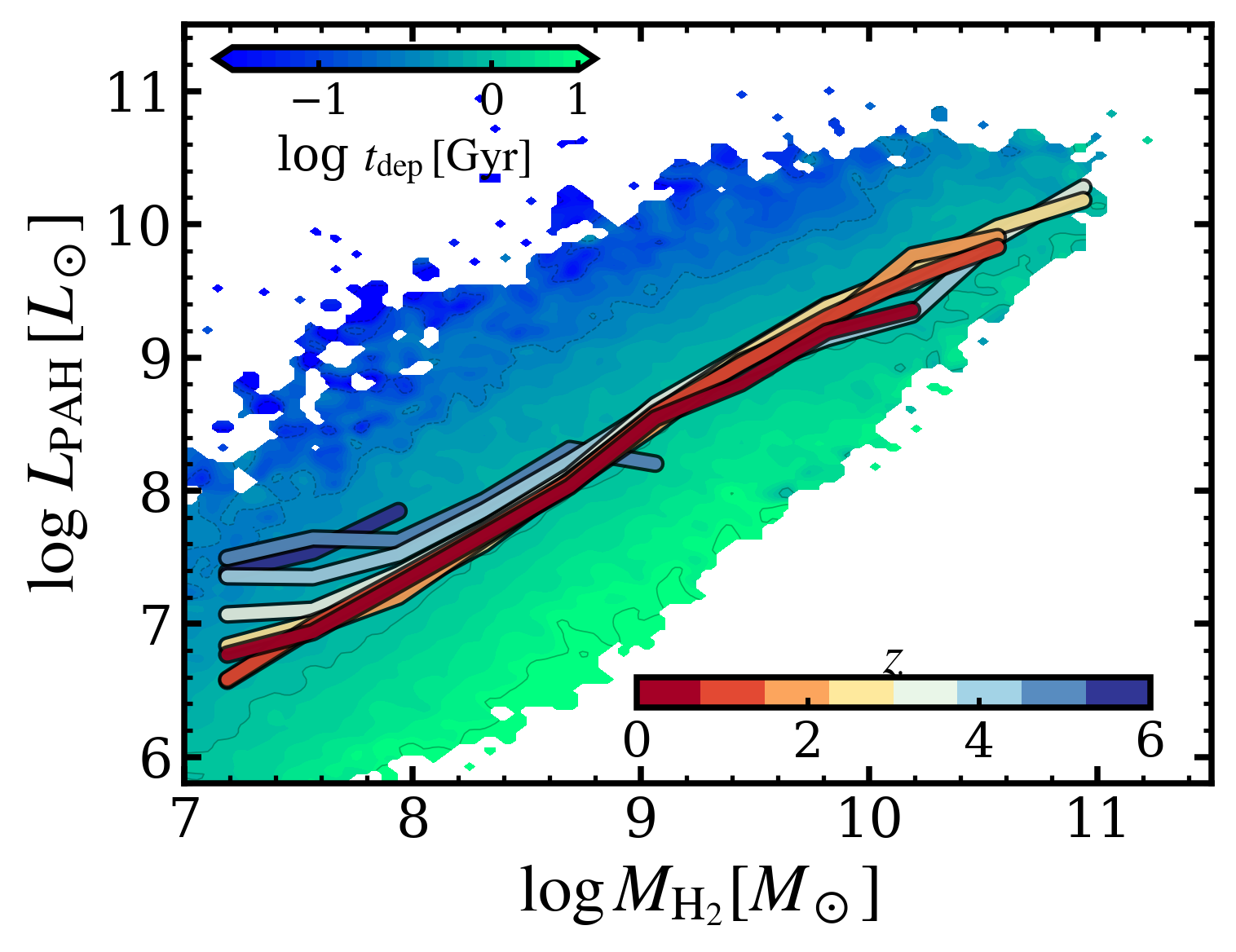}
        \caption{$L_{\rm PAH}$ vs molecular mass}
    \end{subfigure}

    \vspace{0.4cm}

    \begin{subfigure}{0.45\textwidth}
        \centering
        \includegraphics[width=\linewidth]{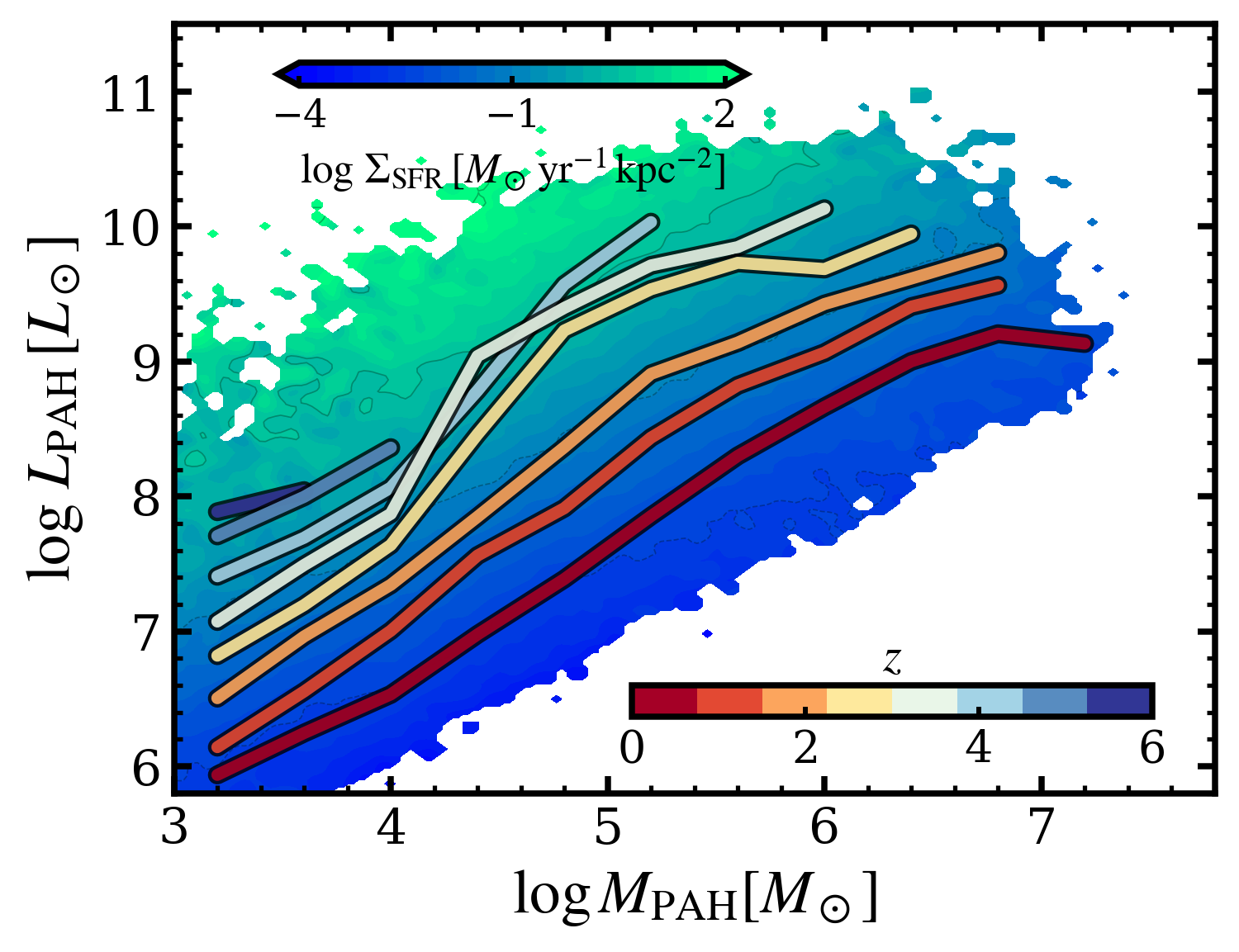}
        \caption{$L_{\rm PAH}$ vs $M_{\rm PAH}$}
    \end{subfigure}
    \hfill
    \begin{subfigure}{0.45\textwidth}
        \centering
        \includegraphics[width=\linewidth]{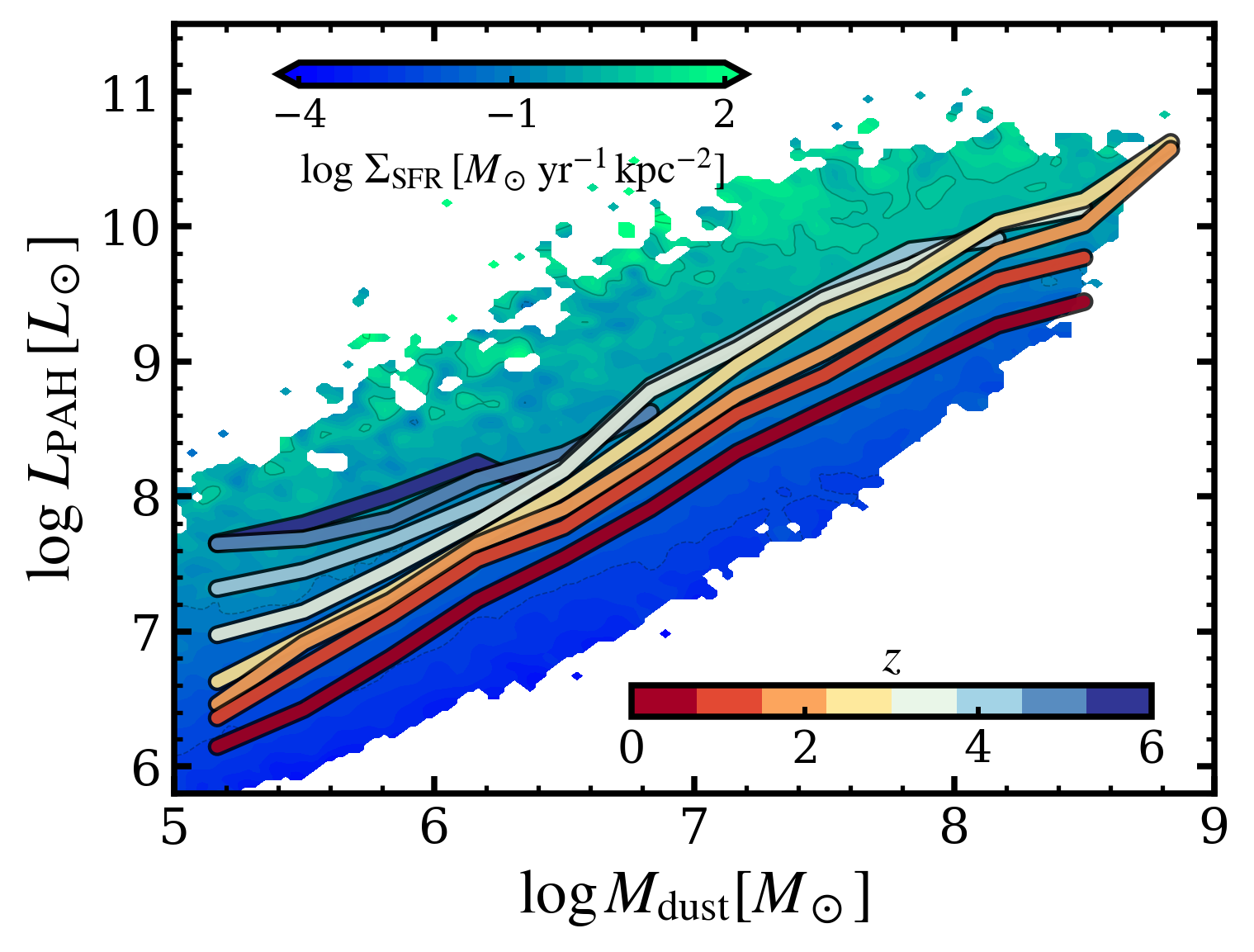}
        \caption{$L_{\rm PAH}$ vs $M_{\rm dust}$}
    \end{subfigure}

    \caption{
    \textbf{PAH luminosity scaling relations.} Total PAH luminosity (all features) as a function of SFR, molecular gas mass $M_{\rm H_2}$, PAH mass $M_{\rm PAH}$, and dust mass $M_{\rm dust}$. Solid lines are median trends for galaxies at different redshift. Contours are color-coded according to their PAH mass fraction $q_{\rm PAH}$ (top left), depletion time $t_{\rm dep}$ (top right) or SFR surface density $\Sigma_{\rm SFR}$ (bottom).}

    \label{fig:PAH_scaling}

\end{figure*}

How well does PAH emission trace fundamental galaxy properties? Thanks to our detailed galaxy evolution framework coupled with RT post-processing, we are able to address this question and provide scaling relations that can inform observations. Figure \ref{fig:PAH_scaling} shows the relation between $L_{\rm PAH}$ (the total luminosity summed over all features) and four key galaxy properties: SFR, molecular gas mass, PAH mass, and dust mass.

PAH emission traces both SFR and molecular gas mass remarkably well, with relatively modest evolution across redshift (top panels of Figure \ref{fig:PAH_scaling}). These variations are however informative, as they shed light on the physics driving the scaling relations. The normalization of the $L_{\rm PAH}-{\rm SFR}$ relation increases slightly with decreasing redshift, reflecting the fact that low-$z$ galaxies are more PAH-rich as a result of the growing importance of grain shattering (Section \ref{sec:pah_abundance}). Indeed, at fixed SFR, more PAH-rich galaxies are more PAH-luminous. Notably, $L_{\rm PAH}$ shows a mild downturn at high SFR that becomes increasingly pronounced at $z \lesssim 2$. This can be attributed to the previously discussed drop in $q_{\rm PAH}$ in highly star-forming (and high-metallicity; see Figure \ref{fig:PZR:main}) systems, where SN destruction of small grains overcomes ISM grain growth. The $L_{\rm PAH}-M_{\rm H_2}$ relation shows even weaker redshift evolution, with only a slightly higher normalization for high-$z$ galaxies. This is due to their shorter depletion times, $t_{\rm dep} = M_{\rm H_2}/{\rm SFR}$: at fixed molecular gas mass, they host more star formation capable of producing UV radiation and thus exciting PAHs, making them brighter. \rev{We highlight here that, since PAHs reside only in the diffuse phase in our model, this correlation does not reflect a spatial association, but arises because $M_{\rm H_2}$ strongly correlates with the SFR (by construction, as it drives star formation in our SAM), and hence with the radiation field illuminating PAHs \citep[see also][]{Nara26}.}

While the role of the radiation field -- so far quantified through the SFR -- is already clear from the above, it becomes central to the relations between PAH luminosity and PAH or dust mass (bottom panels of Figure \ref{fig:PAH_scaling}). Although a clear positive correlation is present at all redshifts, here the redshift evolution is substantially stronger, particularly for $L_{\rm PAH}-M_{\rm PAH}$. At fixed $M_{\rm PAH}$ or $M_{\rm dust}$, galaxies are more PAH-luminous at higher redshift, because of their more intense radiation fields -- here quantified by the SFR surface density $\Sigma_{\rm SFR}$. This result agrees with the findings of \cite{Nara26} in hydrodynamic simulations: PAH luminosity only marginally traces PAH mass, since the light-to-mass ratio depends dramatically on the UV radiation field of the galaxy, and hence, by extension, on redshift. \rev{This has direct implications for observations. First, PAH-poor galaxies at high redshift can be brighter in PAH emission than local PAH-richer objects, making them easier to detect. Second, PAH masses inferred from PAH emission alone can be very misleading and biased with redshift if the role of the radiation field is not taken into account.} We return to this point in Section \ref{sec:overallpicture}.

For reference, we fit the median relation $\log L_{\rm PAH} = a(z) + b(z)\,\log X$ separately in each redshift bin, where $X$ is in turn the SFR, the molecular gas mass, the PAH mass and the dust mass. The best-fit coefficients are listed in Table \ref{tab:pah_fits}.

\begin{deluxetable*}{ccccccccc}
\tabletypesize{\footnotesize}
\tablewidth{0pt}
\tablecolumns{9}
\tablecaption{Linear fits to the binned median $L_{\rm PAH}$ scaling
              relations.\label{tab:pah_fits}}
\tablehead{
  \colhead{} &
  \multicolumn{2}{c}{SFR $[M_\odot\,{\rm yr}^{-1}]$} &
  \multicolumn{2}{c}{$M_{\rm H_2}$ $[M_\odot]$} &
  \multicolumn{2}{c}{$M_{\rm PAH}$ $[M_\odot]$} &
  \multicolumn{2}{c}{$M_{\rm dust}$ $[M_\odot]$} \\[-2pt]
  \cline{2-3} \cline{4-5} \cline{6-7} \cline{8-9} \\[-8pt]
  \colhead{$z$} &
  \colhead{$a$} & \colhead{$b$} &
  \colhead{$a$} & \colhead{$b$} &
  \colhead{$a$} & \colhead{$b$} &
  \colhead{$a$} & \colhead{$b$}
}
\startdata
0.00 & $8.45 \pm 0.02$ & $0.803\,(66)$  & $-0.11 \pm 0.24$ & $0.942\,(30)$  & $3.08 \pm 0.05$ & $0.896\,(10)$  & $0.98 \pm 0.07$ & $1.007\,(12)$  \\
0.51 & $8.47 \pm 0.01$ & $0.864\,(22)$  & $-0.51 \pm 0.10$ & $0.993\,(11)$  & $3.15 \pm 0.06$ & $0.972\,(13)$  & $1.15 \pm 0.08$ & $1.027\,(11)$  \\
1.04 & $8.38 \pm 0.01$ & $1.010\,(44)$  & $-0.87 \pm 0.14$ & $1.030\,(17)$  & $3.60 \pm 0.07$ & $0.957\,(13)$  & $1.23 \pm 0.07$ & $1.037\,(10)$  \\
2.07 & $8.24 \pm 0.01$ & $0.990\,(9)$   & $-0.34 \pm 0.06$ & $0.976\,(7)$   & $4.16 \pm 0.13$ & $0.928\,(31)$  & $1.63 \pm 0.10$ & $1.011\,(14)$  \\
3.11 & $8.13 \pm 0.01$ & $0.916\,(16)$  & $0.36 \pm 0.10$  & $0.903\,(11)$  & $3.98 \pm 0.11$ & $1.061\,(27)$  & $2.35 \pm 0.09$ & $0.922\,(15)$  \\
3.95 & $8.06 \pm 0.02$ & $0.846\,(25)$  & $1.53 \pm 0.28$  & $0.777\,(34)$  & $4.48 \pm 0.21$ & $0.969\,(69)$  & $2.99 \pm 0.10$ & $0.845\,(17)$  \\
5.03 & $8.00 \pm 0.06$ & $0.62\,(12)$   & $3.69 \pm 0.80$  & $0.52\,(10)$   & $5.70 \pm 0.16$ & $0.652\,(53)$  & $5.01 \pm 0.24$ & $0.504\,(48)$  \\
5.92 & $7.91 \pm 0.11$ & $0.63\,(18)$   & $3.18 \pm 1.16$  & $0.58\,(15)$   & $6.73 \pm 0.31$ & $0.35\,(12)$   & $5.77 \pm 0.47$ & $0.37\,(10)$   \\
\enddata
\tablecomments{Linear fits to the binned median relations, $\log L_{\rm PAH} = a(z) + b(z)\,\log X$, performed independently in each redshift bin for $X = {\rm SFR}$, $M_{\rm H_2}$, $M_{\rm PAH}$, and $M_{\rm dust}$, in the units given in the column headings. $L_{\rm PAH}$ is the total PAH luminosity in $L_\odot$. Fits are performed on the median relation rather than on individual galaxies, so that each bin in the independent variable contributes equally regardless of its occupancy. All uncertainties are $1\sigma$ errors obtained by bootstrap resampling of the galaxy sample.  For the slopes $b$, they are given in parentheses in units of the last quoted digit (e.g., $0.803\,(66) \equiv 0.803 \pm 0.066$).}
\end{deluxetable*}

\subsection{Comparison with observations}
\label{sec:obs}
To assess the reliability of the scaling relations predicted by our model, we compare them against observational determinations of the $L_{\rm PAH}-{\rm SFR}$ and $L_{\rm PAH}-M_{\rm H_2}$ relations currently available in the literature.

At low redshift, \citet{Xie19} provide PAH luminosities for a sample of local star-forming galaxies with SFRs calibrated against the $[{\rm Ne\,II}]$ and $[{\rm Ne\,III}]$ lines, while \citet{Shipley16} calibrate PAH luminosities against extinction-corrected ${\rm H}\alpha$-derived SFRs for local star-forming galaxies at $z<0.4$. \citet{Pope08} instead target a more extreme population -- local starbursts and high-redshift submillimeter galaxies (SMGs) -- providing a fit that probes the high-SFR regime. At intermediate redshift, \citet{Cortzen19} compile a large, multi-population sample of local and $z \lesssim 4$ galaxies (5MUSES star-forming galaxies, local ULIRGs, SMGs, and high-$z$ starbursts) with joint PAH and CO measurements, from which they derive a fit that is nearly redshift independent. Finally, two recent JWST/MIRI-based studies extend these relations into the cosmic noon regime previously inaccessible with \textit{Spitzer}: \citet{SB24} use SMILES MIRI photometry to measure PAH luminosities for typical star-forming galaxies at $z\approx 1-3$ with ALMA-derived molecular gas masses, and \citet{McKinney26} present MIRI/LRS spectroscopy of a sample at $0.6 < z < 2.5$, extending the PAH--SFR relation to higher IR luminosities than probed by SMILES alone. \rev{We caution that these observational determinations are not fully homogeneous. While some of the works \citep[e.g.,][]{Shipley16, Cortzen19, McKinney26} measure PAH luminosities through spectral decomposition, the approach closest to the one adopted here, \citet{SB24} derive PAH luminosities from photometric SED fitting, and hence rely on the PAH spectrum assumed in their templates \citep{DraineLi07}.}

\begin{figure*}[]

    \centering
    \includegraphics[width=2\columnwidth]{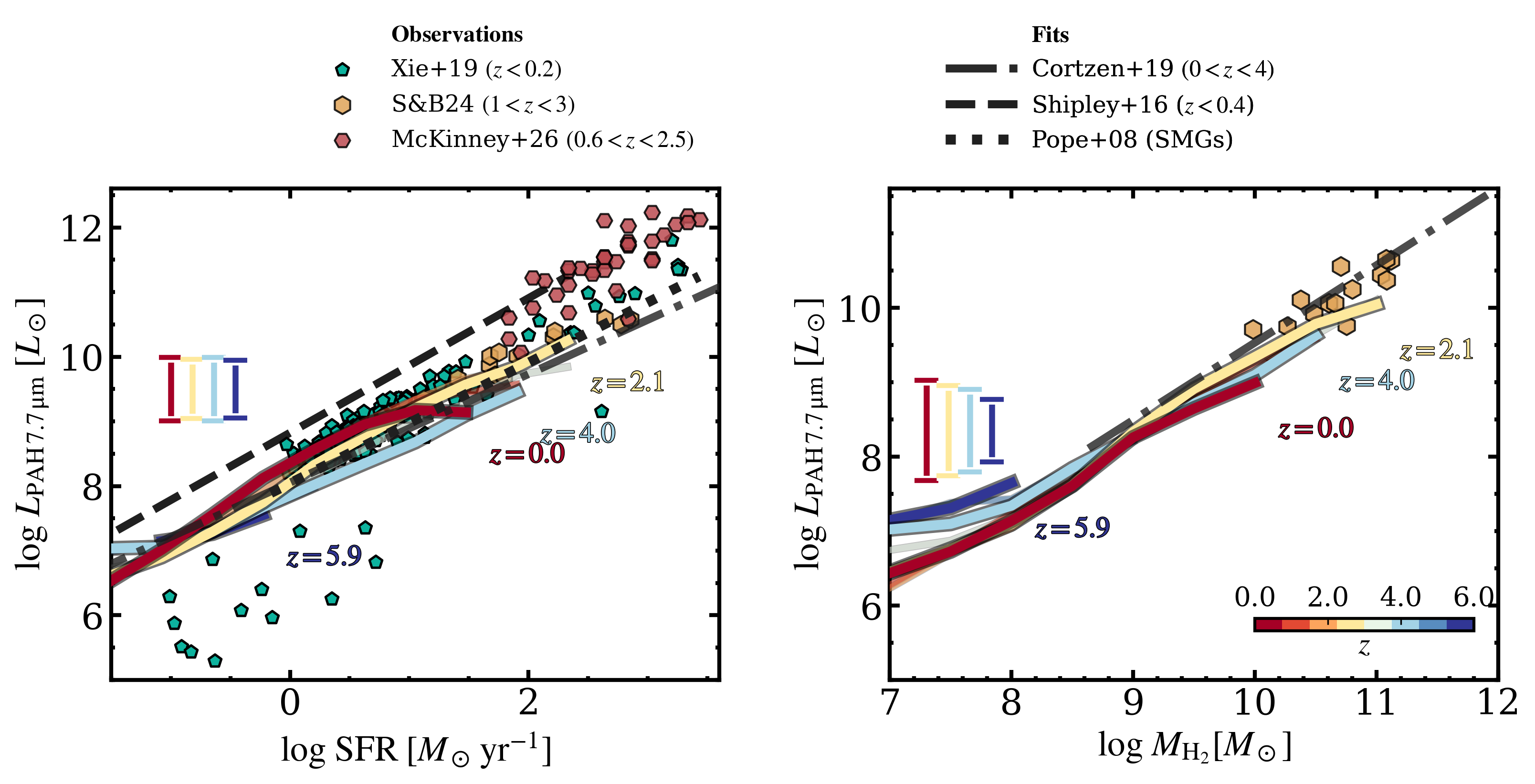}

    \caption{\textbf{Our model well reproduces observations of the PAH $7.7\,\mu{\rm m}$ luminosity vs.\ SFR and molecular mass.} PAH $7.7\,\mu{\rm m}$ luminosity $L_{7.7\,\mu{\rm m}}$ as a function of SFR (left panel) and molecular gas mass (right panel) as predicted by our model at various redshifts, compared to observations. Solid lines represent median trends, with the typical $16$--$84$th percentile dispersion shown as errorbars for selected redshifts. We report observational points from \citet{Xie19,SB24,McKinney26}, as well as observationally derived fits from \citet{Pope08,Shipley16,Cortzen19}.}

    \label{fig:PAH_scaling:obs}
\end{figure*}

Figure \ref{fig:PAH_scaling:obs} compares our model predictions for the $L_{7.7\,\mu{\rm m}}-{\rm SFR}$ (left panel) and $L_{7.7\,\mu{\rm m}}-M_{\rm H_2}$ (right panel) relations against this observational compilation\footnote{Here, in contrast to the rest of the paper, we use the brightest and most commonly observed feature ($7.7\,\mu{\rm m}$) for comparison with the data. In our model, individual feature ratios are not meaningful since we do not track the PAH size distribution: the $7.7\,\mu{\rm m}$ feature simply carries $\approx 50\%$ of the total predicted PAH luminosity.}. 
Overall, we find good agreement between the model and the data: most observed points and fits fall within our $z\approx0-3$ predictions, though we note that observations typically reach higher SFRs than those probed by our simulations. Over the range where they overlap, our predicted slopes compare well with those obtained from fits to the data.
The agreement is particularly good around cosmic noon: our $z\approx 2.1$ relation lies almost on top of the \citet{SB24} determination, which is encouraging given that this is approximately the epoch at which our model predicts the PAH $7.7\,\mu{\rm m}$ emission to peak.

Although less densely sampled by data, the $L_{7.7\,\mu{\rm m}}-M_{\rm H_2}$ relation shows a similarly concordant picture, with our model reproducing both the normalization and slope of the \citet{Cortzen19} fit and the \citet{SB24} points. In particular, the lack of redshift evolution reported in the former work is well reproduced by our model.

This comparison against observations shows that our model captures the main trends found across a varied and inhomogeneous set of PAH observations spanning cosmic time, and it validates -- for the first time from a theoretical point of view -- the empirical relations commonly used to trace star formation and molecular gas mass from PAH emission.


\section{An overall picture for PAH mass and emission evolution}
\label{sec:overallpicture}

\begin{figure}[]

    \centering
    \includegraphics[width=0.99\columnwidth]{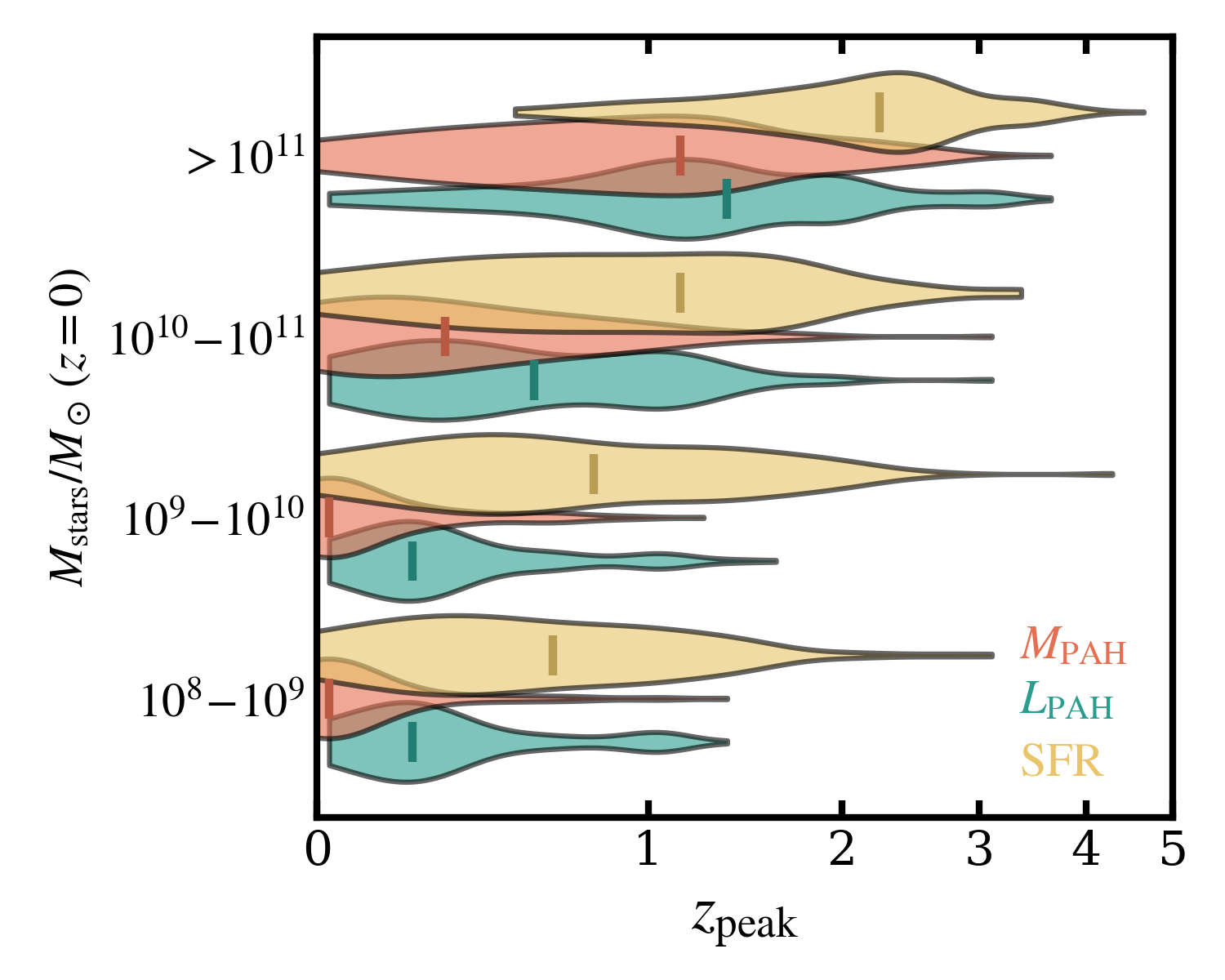}

    \caption{\textbf{PAH emission peaks at a redshift intermediate between the peaks of SFR and PAH abundance.} Distributions of the redshift at which PAH abundance (orange), PAH luminosity (green), and SFR (yellow) each reach their maximum during the evolution of galaxies, split by their $z=0$ stellar mass as indicated on the vertical axis. The median of each distribution is shown as a vertical line within its violin.} 
    \label{fig:zpeak}
\end{figure}

We now provide a general interpretation of these results, focusing on the evolution of PAH luminosity, which is essential for interpreting the observables discussed in Sections \ref{sec:PAHscaling} and \ref{sec:obs}.

The analysis above shows that observed PAH emission depends on, and therefore traces, both the PAH abundance and the intensity of the radiation field. However, we showed in Section \ref{sec:PZR} that in our top-down PAH formation scenario, PAH abundance grows once galaxies move over their star-forming, dense-gas-dominated phase. Since PAH luminosity is effectively a \textit{convolution} of PAH abundance and the radiation field (and hence of star formation), these two competing trends produce a time offset between the epoch of peak PAH brightness and the epoch of peak PAH abundance.

This offset is reported in Figure \ref{fig:zpeak}, which shows, for galaxies in different bins of $z=0$ stellar mass, the distribution of redshifts at which SFR, $L_{\rm PAH}$, and $M_{\rm PAH}$ reach their maximum. As expected, more massive galaxies evolve earlier, reaching the peak of each quantity at higher redshift. In every mass bin the $L_{\rm PAH}$ peak falls systematically between the SFR peak and the $M_{\rm PAH}$ peak. We note that the apparent decline of $M_{\rm PAH}$ at $M_{\rm star} \gtrsim 10^{10}\,M_\odot$ simply reflects gas reduction during the quenching phase of these galaxies, rather than any PAH- or dust-specific process. The PAH mass fraction $q_{\rm PAH}$ indeed continues to rise toward low $z$. Across our sample, the time offset between the $M_{\rm PAH}$ and $L_{\rm PAH}$ peaks is $\approx 0.7$–$2.3$ Gyr, with more massive galaxies showing shorter offsets due to their evolution generally characterized by shorter timescales.

\begin{figure*}[]

    \centering
    \includegraphics[width=2\columnwidth]{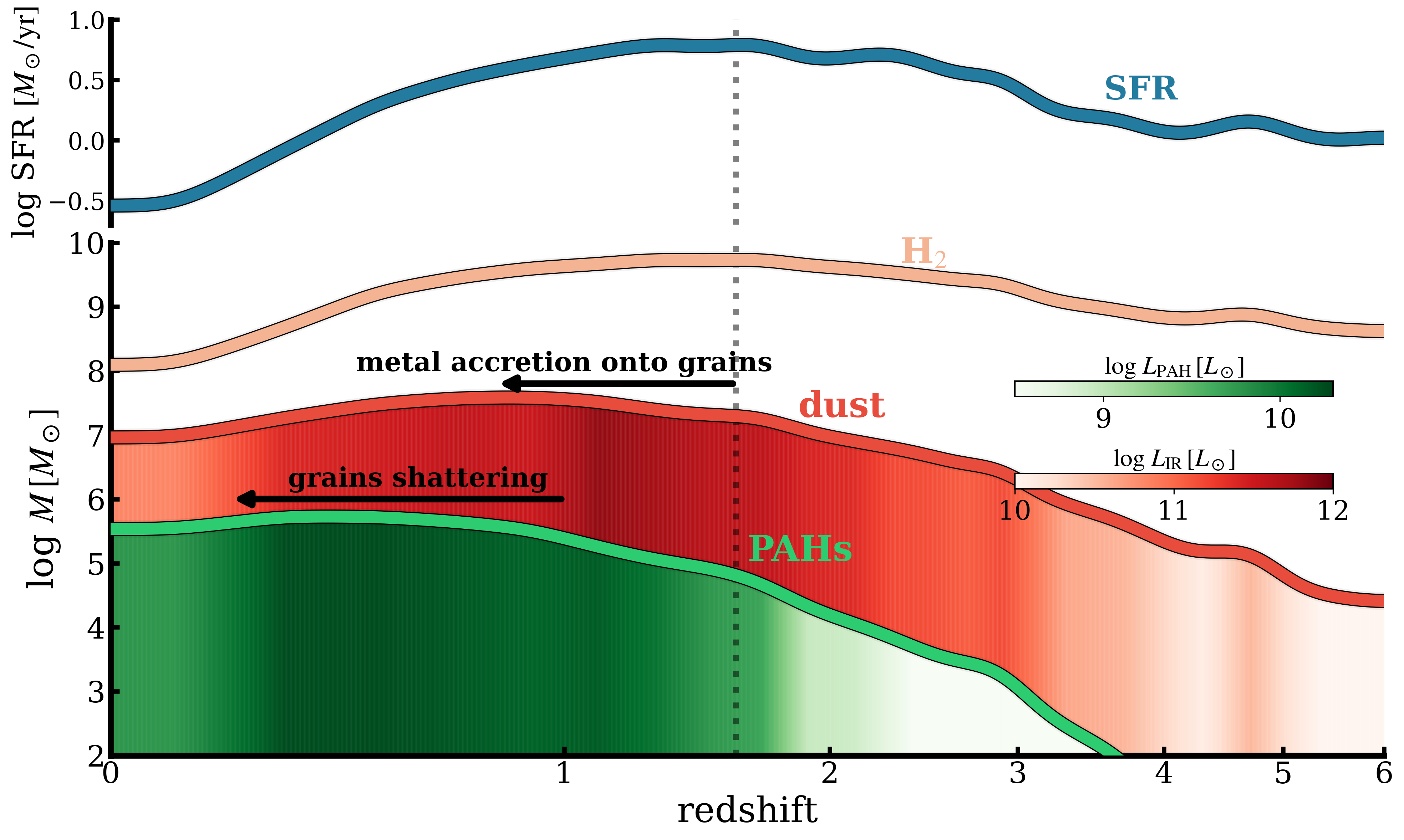}
    \caption{\textbf{Evolution and time shift of key quantities for a sample of galaxies.} The plot shows the redshift evolution of some key quantities for a sample of simulated galaxies with $10^{10} \leq M_{\rm star}/M_\odot < 10^{11}$ -- namely the SFR (top panel), molecular mass, dust mass and PAH mass (bottom panel). Solid lines are medians of the sample. The vertical dotted line marks the redshift of the peak of the SFR. The dust peak is delayed with respect to the SFR and $M_{\rm H_2}$ peak, due to ISM reprocessing and accretion of metals onto grains, and the PAH abundance grows at even later times due to continuous shattering in the diffuse ISM. The area below the dust and PAH curves is colored according to the IR and PAH luminosity, respectively in red and green, which is not monotonic with the abundance.} 
    \label{fig:fullevo}
\end{figure*}

Figure \ref{fig:fullevo} illustrates this offset more explicitly, showing the median evolutionary tracks of several key quantities for simulated galaxies with $z=0$ stellar mass $10^{10} \leq M_{\rm star}/M_\odot < 10^{11}$: SFR, molecular gas mass, dust and PAH abundance, as well as IR and PAH luminosity. Both the SFR and the molecular gas mass peak at $z \approx 1.5$, after which these galaxies gradually quench toward $z=0$.

It is worth noting two features of this figure. First, both the dust and PAH abundances peak \textit{after} the SFR peak. For dust, this delay reflects grain growth in the ISM, which builds most of the dust mass some time after the peak of dust production (and hence after the SFR peak). The delay is more pronounced for PAHs, since in our model PAHs form via grain shattering (Section \ref{sec:PZR}), a process that requires diffuse gas and therefore only becomes efficient once the dense molecular gas reservoir has declined (Section \ref{sec:dust_proc}). This pushes the PAH abundance peak even further away from the SFR peak than the dust peak.
Second, the luminosity of dust and PAHs (red and green shaded regions, respectively) does not simply track their abundance. This is especially clear for PAHs: while $M_{\rm PAH}$ continues to rise and only flattens near $z \approx 0$, the brightest PAH emission actually occurs earlier, at $z \approx 1$ -- close to the peak of SFR and molecular gas -- when the PAH abundance is only about half its maximum. At the same time, $L_{\rm PAH}$ clearly drops at $z \approx 0$, when the PAH abundance is still very high. This behavior directly confirms the convolution between PAH abundance and the radiation field proposed above.


\section{Discussion}
\label{sec:discussion}

Given the novelty of modelling PAHs in galaxy evolution simulations, and since this work is the first to produce large statistics for PAH abundance and observables, it is worth placing our results in the context of alternative interpretations and discussing the main limitations of our approach.

\subsection{The galaxy evolution model}
\label{sec:disc:galevo}
While the SAM framework adopted here allows us to make population-wide predictions across cosmic time, this comes at the expense of ISM detail. As in most SAM treatments of the ISM \citep{Somerville15}, we split the gas into a dense and a diffuse phase with fixed, assumed thermodynamic properties (density and temperature), rather than following the actual dynamical state of the gas -- a strong simplification given how relevant grain collisions are to our results.

Hydrodynamic simulations relax this assumption by computing grain velocities self-consistently from the resolved gas dynamics (e.g., \citealt{McKinnon18, Caleb_prep}). A recent step further in this direction is CALIMA \citep{curro26}, which couples on-the-fly dust and PAH evolution directly to radiative transfer and non-equilibrium thermochemistry in a resolved, multiphase ISM.
Using hydrodynamic simulations \rev{in which, as here, PAHs form via shattering,} \citet{Nara26} also reproduce the observations reasonably well. Since both models rely on the top-down channel, this is not an independent test of the channel itself. Rather, it shows that when PAHs are assumed to form top-down, the resulting predictions are robust to how the ISM is modelled, from a two-phase SAM to resolved hydrodynamics. Consistently, in \citet{Parente26} we explicitly tested the sensitivity of the GSD on the assumed two-phase thermodynamics of the SAM and found only modest changes, supporting the robustness of our results.

\rev{A related limitation of our galaxy evolution model concerns the spatial origin of PAH emission. Recent observations have shown that PAH emission correlates tightly with CO on resolved scales \citep{Leroy23, Chown25}, commonly attributed to photodissociation regions (PDRs) at the surfaces of molecular clouds \citep[e.g.,][]{Chown24}. In our model, PAHs reside only in the diffuse phase, and the $L_{\rm PAH}$--$M_{\rm H_2}$ correlation arises indirectly through the SFR (Section \ref{sec:PAHscaling}). Since PAH luminosity depends on both abundance and illumination, bright PAH emission from PDRs does not necessarily imply that most of the PAH mass resides there. Our model cannot establish whether the observed PAH--CO correlation reflects a physical association, nor where most PAHs reside and where their emission originates. Addressing these questions requires resolved simulations that follow both dust and molecular physics, together with radiative transfer on PDR scales.}

\subsection{The PZR and PSR in the context of alternative explanations}
\label{sec:disc:alternatives}

Several explanations have been proposed in the literature for the PZR and the anti-correlation of the PSR. One of the most common is that photodestruction by UV photons suppresses PAH abundance in highly star-forming and/or low-metallicity (hence poorly shielded by dust) environments \citep[e.g.,][]{Madden06, Hunt10, Xie19, Chastenet25}. Our model does not include this process, yet it reproduces both relations at galaxy-integrated scales (Figures \ref{fig:PZR:main} and \ref{fig:PZR:sSFR}): young, gas-rich galaxies are PAH-poor simply because they have not yet had time to grow and shatter their grains down to PAH sizes (Section \ref{sec:dust_proc}). 
Photodestruction is therefore not required on galaxy-integrated scales, although resolved observations suggest it operates locally, e.g. within H\,\textsc{ii} regions (Section \ref{sec:disc:photoproc}).
Resolved observations of Sextans A and M101 favor \emph{inhibited growth} over enhanced destruction to explain low PAH abundances at low metallicity \citep{Tarantino25, Whitcomb26}.
\rev{Our model shares the idea that growth, rather than destruction, limits PAH abundance. In our model, low PAH abundances at low metallicity arise because ISM processing is slow: metal-poor galaxies have little accretion and little dust available for shattering, so that their dust remains dominated by the large grains and little mass reaches PAH sizes. In \citet{Whitcomb26}, it is instead the growth of PAHs themselves that is suppressed.}

Another process commonly invoked to explain the higher PAH abundances of high-metallicity galaxies is the delayed injection of carbonaceous grains (and PAHs) by AGB stars \citep[][]{Galliano08}. When we switch off carbon grain production by AGB stars in our model (Figure \ref{fig:PZR:colorage}), we recover almost exactly the same PZR. In our model, the delay needed to build up PAHs is supplied by ISM processing itself rather than by the delayed stellar channel. However, this test only concerns the delayed injection of \emph{large} carbonaceous grains. It says nothing about the direct seeding of PAHs by AGB winds, a channel absent from our model.

\subsection{The (missing) bottom-up formation channel}
\label{sec:disc:bottomup}
PAHs may form in two ways: ``top-down'', through the fragmentation of larger carbonaceous grains, or ``bottom-up'', by growing from smaller hydrocarbon precursors, either in the winds of carbon-rich AGB stars or in the ISM itself \citep[e.g.,][]{Frenklach89, Cherchneff92, Reizer22}. \rev{Recent observations support bottom-up formation in the ISM. Individual PAHs have been detected in the dark cloud TMC-1 at unexpectedly high abundances \citep[e.g.,][]{Wenzel25}, and JWST observations of the Orion Bar suggest that similar chemistry may also operate in warmer, UV-irradiated gas \citep{Goicoechea25}.}

Our model, like most current models tracking PAH abundances in galaxies \citep[e.g.,][]{HM20, Narayanan23, Caleb_prep}, includes only the top-down channel. Its success in reproducing the observed scaling relations therefore shows that this channel is \emph{sufficient} at the population level, not that other channels are absent. In our model, PAHs are seeded by shattering and then grow through the same generic accretion physics applied to all carbonaceous grains, with no PAH-specific chemistry. We also assume that PAHs reside only in the diffuse phase, whereas resolved observations of nearby galaxies find higher PAH fractions in molecular regions \citep{Chastenet19, Tarantino25}\rev{, and PAH growth via gas-phase accretion in moderately dense, translucent gas has been invoked to explain the $R_V$--$q_{\rm PAH}$ anti-correlation in the Milky Way \citep{Zhang25}}.

The CALIMA simulation \citep{curro26} takes a first step beyond this picture: PAHs can be directly seeded by AGB winds, and carbon accretion onto existing rings is governed by a chemistry-dependent \textit{accretion probability} (e.g., sensitive to the photostability of the resulting [C-PAH] complex and to hydrogenation state) rather than the standard dust-grain accretion rate.

\subsection{PAH photoprocessing}
\label{sec:disc:photoproc}
Our shattering-driven picture of PAH formation is built mainly on classical grain-grain collision physics \citep{Jones96}, with dust grains and PAHs treated as solid particles that grow and fragment through collisions. What is entirely absent from this picture is photoprocessing -- the destruction and chemical processing of PAHs by the UV photons of the radiation field that excites their emission.
Theoretical and laboratory work \citep[e.g.,][]{Jochims94, Allain96, Montillaud13, Murga19} predicts that photodissociation preferentially destroys the smallest PAHs once they absorb photons energetic enough to exceed their dissociation threshold. This is a completely different destruction channel from the mechanical shattering/coagulation cycle that governs our GSD, and it specifically affects the size range we identify with PAHs. Moreover, photoprocessing also drives the hydrogenation and aliphatic-to-aromatic conversion of PAHs \citep[e.g.,][]{Jones13, Murga19} -- the same physical process our model already assumes, but treats as instantaneous and phase-dependent only, rather than as radiation-driven.

Similarly, a missing piece of physics in our model is photoionization. We assign the PAH ionization fraction as a fixed function of grain size alone \citep[Equation \ref{eq:fion};][]{Draine21, Hensley_astrodust}, independent of the local radiation field intensity or hardness and of the local electron density -- even though, physically, the ionization balance is set by the competition between UV photoionization and electron recombination \citep{Weingartner01}, and can be computed self-consistently once the local ISM physics is known, as recently attempted by \cite{curro26}.

This physics has particularly important consequences for predicted PAH band ratios, since ionized (neutral) PAHs emit preferentially in the $6.2, \, 7.7, \, {\rm and}\,8.6 \, \mu{\rm m}$ ($3.3 \, {\rm and}\,11.3 \, \mu{\rm m}$) bands \citep{DraineLi07} and size-dependent photodestruction is as well traced by PAH band ratios sensitive to molecular size \citep[e.g.,][]{Maragkoudakis20}.
Spatially resolved studies of photodissociation regions find systematic variations in PAH size with local radiation field conditions, generally consistent with preferential destruction of the smallest PAHs in the most intensely irradiated gas \citep[e.g.,][]{Croiset16, Knight21, Maragkoudakis26}.
Also, there is evidence of an overall deficit in total PAH abundance within H\,{\sc ii} regions \citep{Egorov23, Chastenet19, Sutter24, Murgia26}. Such size-sensitive PAH ratio variations are now being measured at cosmic noon as well: spatially resolved JWST/MIRI MRS spectroscopy of $z\sim1$ star-forming galaxies from the PAHSPECS program finds PAH ratios that vary systematically with the hardness of the local UV field, consistent with photo-destruction of small and ionized PAHs shaping the emitted spectrum \citep{Donnan26}.

This growing evidence for the relevance of photoprocessing -- both destructive and ionizing -- to PAH abundance, size distribution, and emitted spectrum makes its self-consistent inclusion a natural next step for our model, requiring spatially and spectrally resolved knowledge of the local radiation field around dust grains and PAHs.
Since this physics is not yet part of our framework, we stress that our results should not be trusted at the level of the detailed internal PAH size distribution or ionization state, but only as an estimate of the total PAH mass and luminosity.

\section{Conclusion}
\label{sec:conclusion}
PAH emission is a powerful and still largely unexplored way of studying galaxy evolution, despite the rapidly growing number of PAH observations from the local Universe out to cosmic noon.

In this work we have presented the first large-volume cosmological simulation of galaxy evolution that combines a physically motivated, grain-size-resolved model of dust \citep{Parente26} and PAH abundance with full radiative transfer post-processing, providing predictions for both PAH abundance and PAH emission over a wide range of cosmic times ($0 < z \lesssim 6$). PAHs are identified with the smallest carbonaceous grains ($a < 15\,\text{\AA}$; $N_{\rm C} \approx 1000$) residing in the diffuse phase of the ISM, and their abundance is regulated self-consistently by the same physics that shapes the whole grain size distribution -- stellar production of large grains, accretion of gas-phase metals, shattering and coagulation in grain--grain collisions, and destruction by SN shocks and thermal sputtering. The resulting SEDs, and hence the mid-IR PAH features, are computed with \textsc{GRASIL} \citep{Silva98}, which treats PAH heating in the single-photon limit, and are decomposed with PAHFIT \citep{Smith07} exactly as done for observed spectra.

Our main results follow.

\begin{itemize}
    \item \textbf{The PAH-metallicity relation --} Our model naturally reproduces the observed PAH--metallicity relation (PZR), with $q_{\rm PAH}$ increasing with ISM metallicity and with a normalization that rises towards low redshift, in good agreement with both local (\citealt{RR15}; \citealt{Aniano20}) and cosmic noon (\citealt{Shivaei24}) determinations. At $z \lesssim 1$ the relation turns over at high metallicity, where SN destruction -- more effective on small grains -- overcomes the saturated accretion of metals onto grains.
    The model also reproduces the observed anti-correlation between $q_{\rm PAH}$ and specific SFR (\citealt{Shim23}; \citealt{Chastenet25}). At fixed metallicity, galaxies are more PAH-rich when they are older and less star forming.

    \item \textbf{A top-down formation scenario --} PAH build-up in our simulation follows a top-down route, with two distinct regimes. At high redshift, when galaxies are dense-gas rich, PAH abundance is mostly set by carbon accretion onto PAH-sized grains. At $z \lesssim 1$, once the ISM becomes diffuse-phase dominated, shattering takes over and keeps increasing $q_{\rm PAH}$ even while the total dust mass slowly declines. Shattering remains essential at all epochs, since in our framework stars only produce large grains and shattering is what generates the first PAH seeds. Notably, this route alone is sufficient: both the PZR and the $q_{\rm PAH}$--sSFR anti-correlation emerge \emph{without} invoking PAH photodestruction in strongly irradiated or poorly shielded environments, and without any direct injection of carbonaceous grains by AGB stars. Neither mechanism is therefore \emph{required} to reproduce these trends at galaxy-integrated scales, although this does not exclude that they operate on resolved scales.

    \item \textbf{PAH luminosity and mass fraction --} The PAH luminosity fraction $L_{\rm PAH}/L_{\rm IR}$ correlates well with $q_{\rm PAH}$, with a scatter of up to $\gtrsim 1$ dex driven by the competition for UV photons between PAHs and the other small grains: at fixed $q_{\rm PAH}$, galaxies with a larger small-to-large grain ratio are fainter in the PAH features. Adding the sSFR removes most of this scatter, and we provide a simple calibration (Equation \ref{eq:qpah_fit}) recovering $q_{\rm PAH}$ from $L_{\rm PAH}$, $L_{\rm IR}$, and sSFR to $\sigma_{q_{\rm PAH}} = 0.12$ dex, i.e., a factor of $\approx 1.3$. This can be used in observational studies as an alternative to traditional SED fitting methods.

    \item \textbf{PAHs tracing galaxy properties --} PAH luminosity traces SFR and molecular gas mass remarkably well and with weak redshift evolution, in good agreement with the available observational determinations -- particularly around cosmic noon, providing theoretical support to the empirical use of PAH features as tracers of star formation and molecular gas. The $L_{\rm PAH}$ relations with PAH or dust mass, in contrast, evolve strongly with redshift: at fixed $M_{\rm PAH}$, galaxies are more PAH-luminous at high $z$ because of their more intense radiation fields, so PAH emission is a poor tracer of PAH mass unless the radiation field is taken into account.

    \item \textbf{PAH luminosity and abundance offset --} Since PAH luminosity is a convolution of PAH abundance and radiation field, PAH emission peaks at a redshift intermediate between the peak of the SFR and the peak of the PAH mass, with a typical offset of $\approx 0.7$--$2.3$ Gyr between the $L_{\rm PAH}$ and $M_{\rm PAH}$ maxima. \rev{This is a genuine prediction that can be tested observationally by comparing the redshift evolution of PAH luminosity and PAH abundance across galaxy populations with JWST and future PRIMA surveys.}
\end{itemize}

Our model of PAH formation remains admittedly simple, and still misses relevant physics such as the bottom-up formation channel and photo-processing by UV photons (Section \ref{sec:discussion}). Even so, this work represents a first step towards fully exploiting PAH observations as a tool for studying galaxy evolution, and towards providing them with a physical foundation from a galaxy evolution perspective. The framework presented here delivers the statistical, cosmologically representative population needed to interpret the PAH scaling relations now measured with JWST/MIRI, and extends these predictions into the high-redshift regime that future facilities such as PRIMA will probe. It is this joint theoretical and observational effort that will turn PAHs into a new and powerful complementary probe of galaxy evolution.\\

\begin{acknowledgments}
This work was funded by NASA ATP programs 80NSSC22K0716 (PI: PT) and 80NSSC24K1223 (PI: DN). This research was carried out in part at the Jet Propulsion Laboratory, California Institute of Technology, under a contract with the National Aeronautics and Space Administration (80NM0018D0004).
\end{acknowledgments}

\bibliography{bib_PAH}{}
\bibliographystyle{aasjournalv7}

\appendix
\section{Impact of the non-PAH grain size distribution on PAH emission}
\label{app:SEDexp}

\begin{figure}[]

    \centering
    \includegraphics[width=0.7\columnwidth]{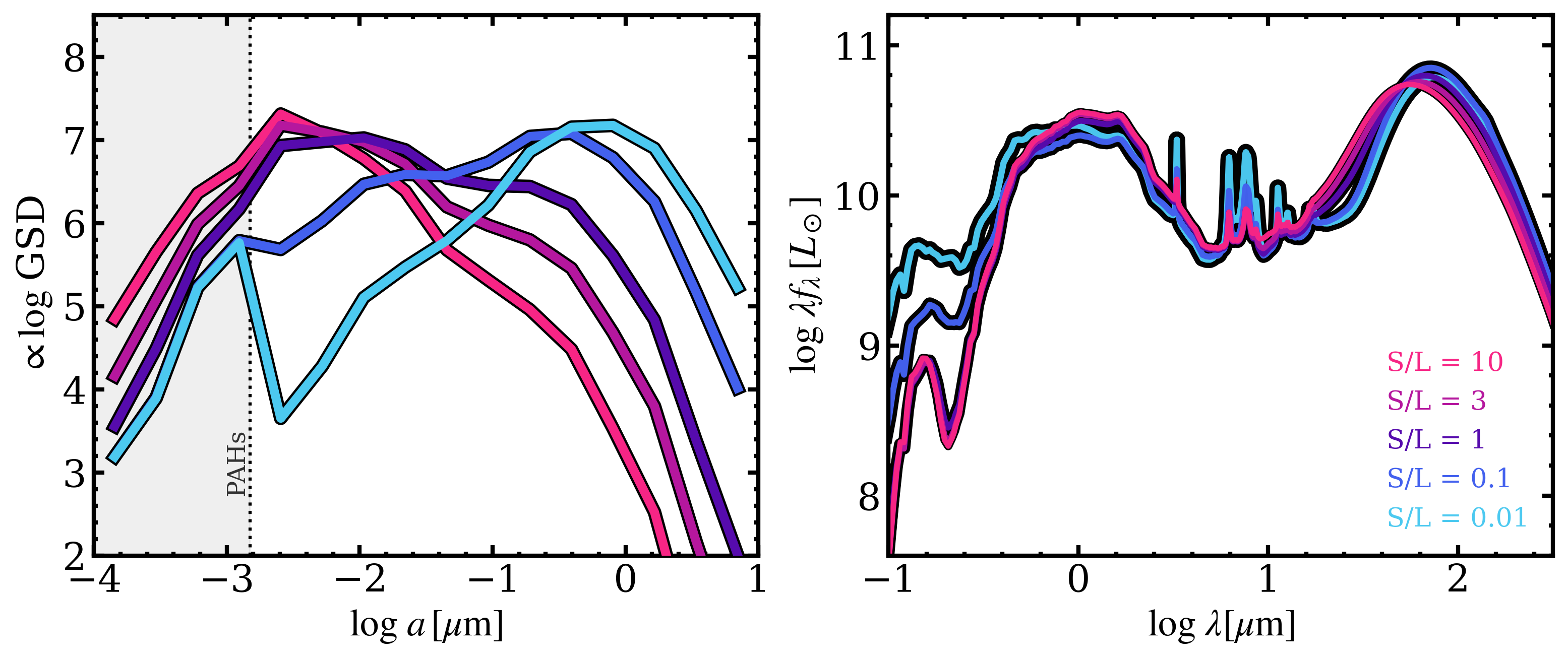}

    \caption{\textbf{PAH emission dependence on the dust grain size distribution.} \textit{Left:} total grain size distribution (carbonaceous $+$ silicate) for five values of the small-to-large (S/L) ratio. The shaded region denotes PAH-sized carbonaceous grains, whose mass is held fixed across experiments. \textit{Right:} corresponding SEDs, colored by the same S/L values. Increasing the abundance of small grains enhances UV attenuation and shifts the IR emission peak to shorter wavelengths, while leaving the total IR luminosity largely unchanged.} 
    \label{fig:fakeSLexperiments}
\end{figure}

Here we present a set of controlled numerical experiments designed to isolate the impact of the grain size distribution on the SED and PAH emission of galaxies. Taking a galaxy at $z=1.04$, we manually vary the small-to-large grain ratio (S/L), with the transition set at $a=0.015\,\mu\mathrm{m}$, while holding fixed both the total dust mass and the mass of carbonaceous grains at PAH sizes ($a<15\,\text{\AA}$), i.e., keeping $q_{\rm PAH}$ constant across all realizations\footnote{\rev{The total GSD can vary at PAH sizes due to small silicate grains, while the carbonaceous PAH mass is kept fixed.}}. The left panel of Figure \ref{fig:fakeSLexperiments} shows the resulting grain size distributions, and the right panel shows the corresponding SEDs. As the abundance of small grains increases, UV attenuation in the SED grows, i.e., the attenuation curve gets steeper\rev{, as shown by the emerging stellar emission at $\lambda \lesssim 1\,\mu$m in the right panel of Figure \ref{fig:fakeSLexperiments}, which decreases with increasing S/L, more strongly at shorter wavelengths}. Small grains absorb UV photons more efficiently and are heated to higher temperatures, as suggested by the blueward shift of the IR emission peak, while the total IR luminosity is almost unaffected. This enhanced absorption by small grains comes at the expense of PAHs, which compete for the same UV photons. With less energy available to absorb, PAHs radiate less and produce shallower mid-IR features. This mechanism explains why, at fixed $q_{\rm PAH}$, galaxies whose grain population is dominated by larger grains exhibit a higher $L_{\rm PAH}/L_{\rm IR}$, since larger grains leave more UV photons (i.e., energy) available for PAHs to absorb.



\end{document}